# Encapsulation of fragrances and oils by core-shell structures from silica nanoparticles, surfactant and polymer: Effect of particle size

Tatiana G. Slavova, Gergana M. Radulova, Peter A. Kralchevsky*, Krassimir D. Danov

*Department of Chemical and Pharmaceutical Engineering, Faculty of Chemistry and Pharmacy, Sofia University, Sofia 1164, Bulgaria*

ORCID Identifiers: Peter A. Kralchevsky: 0000-0003-3942-1411 ; Krassimir D. Danov: 0000-0002-9563-0974

**Abstract**

Oils and fragrances can be encapsulated by using composite shells of silica nanoparticles, polymer and surfactant (potassium oleate). The template for the creation of the core-shell structure is a particle stabilized (Pickering) emulsion. The surfactant adsorbs on the nanoparticles and leads to their reversible hydrophobization and adsorption on the oil/water interface. The outer layer of the self-assembled shell represents a layer from crosslinked polymer. The procedure of encapsulation is simple and includes single homogenization by ultrasound of the formulation that contains all ingredients together. The produced capsules have mean radius in the range between 2 and 11 microns. By order of magnitude and trend, the capsule size follows the law of limited coalescence with respect to the dependence on nanoparticle size and concentration. The composite structure of the shells leads also to dependence on the concentrations of added polymer and surfactant. The produced microcapsules are stable when rinsed with pure water of pH in the range 3 – 10. However, if dispersed in water of pH > 11, the microcapsules are destabilized and release their cargo, i.e., they are pH-responsive. Various fragrances and oils, such as limonene, citronellol, benzyl acetate, and sunflower seed oil were encapsulated. The developed methodology could find applications in any field, in which reversible encapsulation of oily substances is needed.

Keywords:
Encapsulation; Fragrances; Silica nanoparticles; Colloidosomes; Core-shell structures; Pickering emulsions.

___

* Corresponding author. Fax: +359 2 9625643.
  Email address: pk@lcpe.uni-sofia.bg (P.A. Kralchevsky)

# 1. Introduction

In a previous study [1], we developed pH-responsive colloidosome-based microcapsules loaded by oils, including fragrances. The passive release of the encapsulated material through the interparticle openings of the colloidosomes was blocked by a mixed adsorption layer of polymer and surfactant (brick-and-mortar structure). Upon shelf-storage, the obtained microcapsules (in aqueous medium) were stable for at least 8 months. However, upon raising the pH above 6, these capsules were quickly destroyed and released their cargo. Here, we make the next step in this study replacing the 3.5 μm sized silica particles used in Ref. [1] with silica nanoparticles of much smaller diameter, ~ 10 nm, and investigate the structural and functional changes due to the differences in particle size.

Comprehensive reviews on the different colloidal carriers by their chemical nature, morphology and mechanisms of cargo release can be found in Refs. [1-4]. Here, we focus on emulsion-based capsules, which are the subject of the present study. For these colloidal carriers, the cargo is located in the disperse phase, i.e. in the droplets, which are stabilized against coalescence by appropriate amphiphilic molecules, which can be surfactants, polymers, proteins and colloid particles [4-6]. The layer-by-layer technique [7] was applied as a methodology for producing capsules based on emulsion drops that are covered by multilayered polyelectrolyte shells composed of layers from anionic and cationic polymers [8-11]. Particle stabilized emulsions [12-16] and colloidosomes [17-19] have been also used as templates for capsule production [4,20-22]. Another type of template is provided by the mesoporous silica [23,24].

In our previous study [1] we used non-porous hydrophilic silica particles of micrometer size, which were partially hydrophobized by the adsorption of potassium oleate (KOleate), so that they were able to stabilize Pickering emulsions. In contrast with the silanizing agents, such as hexamethyldisilazane (HMDS) and dichlorodimethylsilane (DCDMS), the hydrophobization with KOleate has reversible character. At lower pH values, the surfactant is adsorbed on the particles, renders them hydrophobic and the Pickering emulsion is stable. At higher pH, the KOleate desorbs, the silica particles recover their hydrophilicity and detach from the drops; the microcapsules are destroyed and their cargo is released. In this way, the hydrophobization with KOleate makes the microcapsules pH responsive.

Our goal in the present article is to further extend the methodology from Ref. [1] by replacement of the microparticles with nanoparticles in order to obtain capsules that are (i) stable upon rinsing and do not release the encapsulated material; (ii) monodisperse and visible by optical microscope; (iii) easy to separate from the rest of the formulation and (iv) applicable for encapsulation of various oils and fragrances.

The structure of the paper is as follows. Section 2 describes the used materials and methods. Section 3 summarizes some basic equations used for quantitative analysis of the experimental data. In Section 4, the roles of the basic ingredients, viz. particles, carboxylate and polymer, are investigated and the composition of the formulation is optimized and applied for encapsulation of fragrances and oils. In Section 5, the relation between the sizes of capsules and nanoparticles is analyzed on the basis of the limited coalescence law. Finally, in Section 6 the stability of the produced capsules against the rise of pH and treatment with anionic detergent is investigated. Additional experimental results are presented in the Appendix as a Supplementary Information. In the paper, special attention is paid to the similarities and dissimilarities between the properties of capsules stabilized with microparticles, Ref. [1], and nanoparticles (the present study).

**2. Materials and methods**

*2.1. Materials*

Table 1. Data for the used colloidal $SiO_2$ particles

| $SiO_2$ particles | Specific surface area, $A_S$ (m$^2$/g) | Mean particle radius, $a_{32}$ (nm) |
|---|---|---|
| Ludox SM | 348 ± 30 | 3.9 ± 0.3 |
| Ludox HS-30 | 221 ± 30 | 6.2 ± 0.8 |
| Ludox TM-50 | 130 ± 20 | 10.5 ± 1.6 |
| Aerosil 300 | 300 ± 30 | 4.6 ± 0.5* |

*Radius of the primary particles in fractal-like agglomerates

The spherical $SiO_2$ nanoparticles (colloidal silica) used in the present study are Ludox® SM, Ludox® HS-30 and Ludox® TM-50, products of Sigma-Aldrich (St. Louis, USA). They are supplied as water dispersions containing discrete, hydrophilic, nonporous, spherical particles made of amorphous silica of mass density $\rho_p$ = 2.196 g.cm$^3$. Table 1 shows the specific surface area, $A_S$, of the respective particles as given by the producer and the volume/surface mean particle radius, $a_{32}$, calculated by us (see Section 3). The biggest particles are those of Ludox TM-50 with $a_{32}$ = 10.5 nm. The surface electric charge of the $SiO_2$ particles depends on the pH of the aqueous phase – see Fig. 1a, where data for the zeta-potential of the Ludox TM-50 particles are shown.

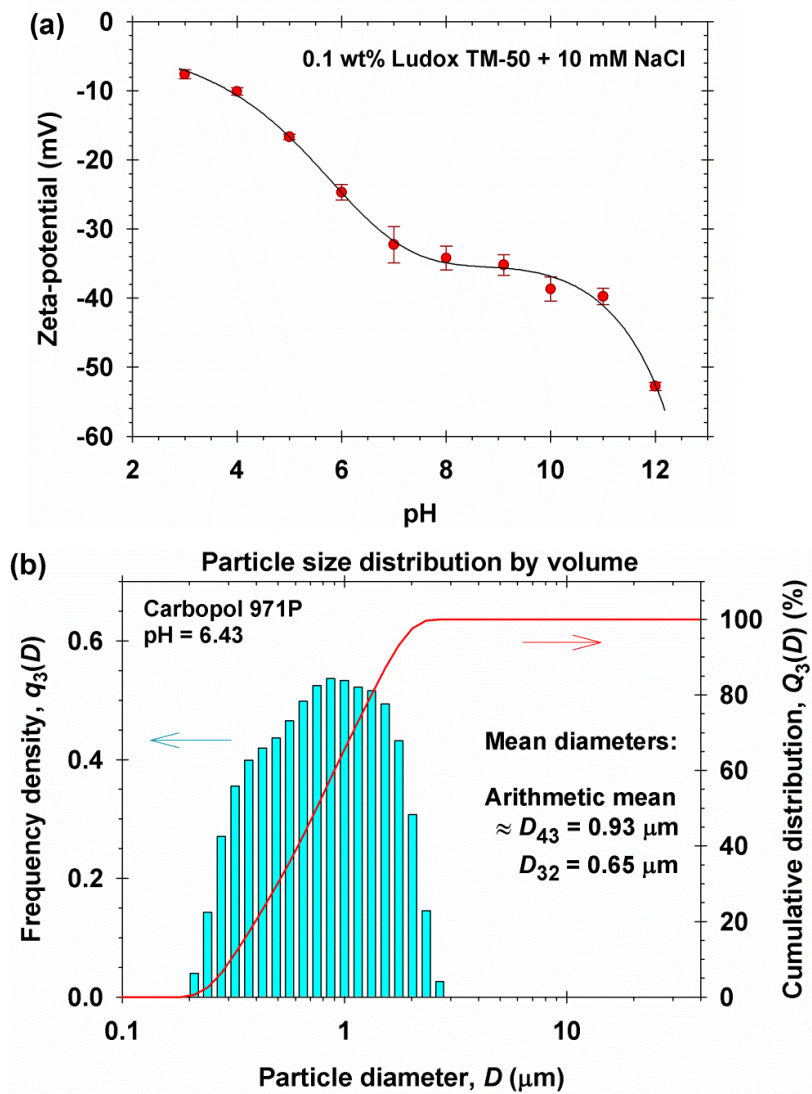

**Fig. 1.** (a) Plot of zeta-potential vs. pH for Ludox TM-50 silica particles in the presence of 10 mM NaCl. (b) Size distribution of the aggregates in an aqueous solution of the polymer Carbopol 971P.

In our experiments, we used also agglomerated fumed silica particles Aerosil® 300 with a specific surface area of 300 m²/g, produced by Evonik Co. (Parsippany, NJ, USA). Aerosil® 300 is supplied as dry powder from fractal-like aggregates, which are composed from adherent primary SiO$_2$ particles. From the specific surface area, one can estimate that the mean radius of these primary particles is $a_{32}$ = 4.6 nm; see Table 1 and Section 3. The average size of the agglomerates of Aerosil® 300 is ca. 2 μm. These agglomerates are fragile. The ultrasound treatment (sonication) used in our emulsification experiments led to breakage of the agglomerates to smaller aggregates of diameter ≈180 nm, as established by us by dynamic light scattering.

The organic liquids subjected to encapsulation were the fragrances limonene, citronellol and benzyl salicylate (products of Sigma Aldrich), and benzyl acetate (product of TCI Co., Tokyo, Japan). Limonene is used not only as fragrance, but also in the agro-food industry as antimicrobial, herbicidal and antioxidant agent [25]. As a representative of the triglycerides, we used sunflower seed oil (SFO) from a local producer. SFO is composed of triglycerides derived from 52 wt% linoleic acid (polyunsaturated omega-6); 34% oleic acid (monounsaturated omega-9), and 12 wt% palmitic acid (saturated) [26]. The SFO was purified from other admixtures by passing through a column filled with Silica Gel and Florisil adsorbent. The measured value of the interfacial tension against pure water was 31 mN/m, a typical value for pure SFO [27], and it did not decrease by more than 0.2 mN/m within 60 min. The other oils were used without any further purification. For better visualization, in most experiments we dissolved 0.12 wt% of the dye Sudan III (Sigma Aldrich) in the oily phase. Thus, the oil drops acquire a reddish color and were better visible under microscope.

We used also sodium chloride (NaCl, product of Sigma Aldrich) and carboxylate – potassium oleate (KOleate) product of Fluka Chemie (Buchs, Switzerland). The KOleate has the property to adsorb with its headgroup to the silica particles and to hydrophobize them [1]. In the experiments on microcapsule destruction, we used the detergent sodium laurylethersulfate with one ethylene oxide group (SLES-1EO, product of Stepan Co., Northfield, IL, USA).

The used polymer was Carbopol® 971P NF, a representative of the Carbopol® homopolymers produced by Lubrizol Co. (Wickliffe, Ohio, USA) that represent polyacrylic acid crosslinked with allyl sucrose or allyl pentaerythritol. In our previous study [1], we tested various polymers and the most stable microcapsules were obtained with Carbopol 971P. The content of carboxylic acid groups in this polymer is 56 – 68 wt%. It is suppled as a solid white powder. In aqueous solutions, the crosslinked Carbopols form microgel particles, which could further aggregate in bigger agglomerates; see Ref. [28], where electronic micrographs of Carbopol aggregates can be seen. Fig. 1b shows the size distribution of the polymer aggregates in a solution of the used Carbopol 971P. The measured zeta-potential of the aggregates of Carbopol® 971P NF is $\zeta$ = -73 mV (0.1 wt% polymer + 10 mM NaCl at pH 7.5) – see Ref. [1], Table A1 therein.

All aqueous solutions were prepared with deionized water (Elix purification system, Millipore). In some experiments, the pH was decreased to 3 by the addition of HCl, or increased to 12.5 by the addition of NaOH. All experiments were carried out at a room temperature of 25 °C.

2.2. Methods

*Preparation of the dispersions.* We used the procedure developed in Ref. [1], which is as follows. First, the silica particles (as water suspension or dry powder) were placed in the working vial and aqueous solution of KOleate was added. The vial was shaken by hand to disperse the particles. Next, water solution of polymer and inorganic electrolyte (NaCl) was added and the vial was shaken again to mix the ingredients. Finally, the oily phase is added. The total volume of the working samples was 10 mL.

To prepare the dispersion, the vial was placed in an ultrasonic pulse homogenizer SKL1500 – IIDN. The sonication was conducted with a sonotrode of diameter 3.24 mm by 1 s long pulses of power output ≈ 200 W, followed by 0.5 s off time. The total sonication time was 10 minutes.

*Rinsing the microcapsules with water*. After the sonication, the vial with the dispersion was left for ≥ 30 min at rest. During this time, gravitational separation of the microcapsules takes place in the form of cream at the top of the water phase (Fig. 2) or as a sediment at the bottom, depending on whether the oil is lighter than water (e.g. limonene) or heavier than water (e.g. benzyl salicylate).

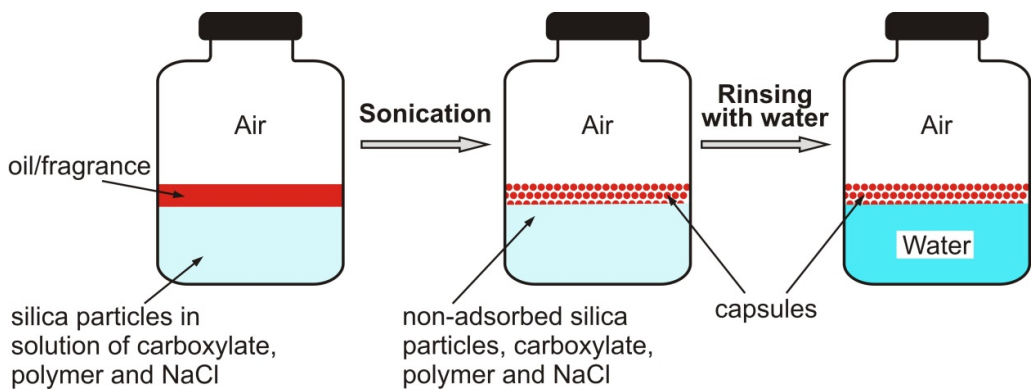

**Fig. 2.** Scheme of the processes of microcapsule preparation and rinsing; details in the text.

Next, the water phase (which may contain nanoparticles, polymer, carboxylate and salt) was removed using a syringe, and replaced with pure water – this is the rinsing step shown in Fig. 2. Next, the vial was shaken by hand. After 30 min at rest, the rinsing was repeated – the water phase was removed again, replaced with water and shaken by hand. The capsules, which survive the described rinsing procedure, will be called *stable capsules*. In the case of unstable capsules, the rinsing leads to release of the encapsulated cargo as a top layer of oil.

*Microscope observations.* To observe the structure of the cream obtained after the sonication and rinsing, the obtained dispersions were studied by microscope. A small portion of the cream was taken by a pipette and placed on a microscope slide, which was after that covered with a glass slip and observed in transmitted light. We used Axioplan microscope (Zeiss, Germany), equipped with Epiplan objectives (10×, 20× and 50×) and connected to CCD camera with digital recorder. To determine the average diameter of the produced microcapsules, the diameters of at least 500 capsules were measured (one by one, semi-automatically) by using image analysis. To ensure correct statistics, we processed video frames corresponding to different positions of the focal plane inside the dispersion.

*Zeta-potentials and particle size distributions*. Zeta-potentials were measured with aqueous suspensions of the used silica particles at various pH values (Fig. 1a). For this goal, the apparatus Zetasizer Nano ZS (Malvern Instruments, UK) was used. This apparatus was used also to measure the mean size of the agglomerates of Aerosil® 300 after the sonication of their suspension. Particle size distributions in solutions of the polymer Carbopol 971P were measured by a laser particle sizer Analysette 22 NanoTec (Fritsch, Germany) with a measuring range 0.01 – 2100 μm; see Fig. 1b.

*Capsule destruction and cargo release*. As in Ref. [1], the capsules were destroyed at high pH or upon the addition of detergent, SLES-1EO, at a sufficiently high concentration. For this reason, after the separation of the capsules at the top of the aqueous phase (Fig. 2), the aqueous liqueur is removed by sucking with a syringe and replaced with another water phase of the same volume and of given pH or SLES-1EO concentration. Then, the vial is shaken by hand. If the capsules are not destroyed, the newly added water phase remains clear. If the capsules are destroyed, the aqueous phase becomes turbid and acquires reddish color (the oil phase has been colored with the red dye Sudan III), which is due to release of the encapsulated material (cargo release) in the form of small emulsion drops; see Section 6 for more details.

## 3. Theoretical background

In Fig. 1b, the quantity $q_3(D)dD$ represents the volume fraction of particles (droplets) with diameters in the interval from $D$ to $D+dD$; $q_3(D)$ is the particle size frequency distribution. Correspondingly, the quantity

$$Q_3(D) = \int_0^D q_3(\hat{D})d\hat{D} \tag{1}$$

is the cumulative particle size distribution; $\hat{D}$ is a variable of integration. The mean diameters $D_{32}$ and $D_{43}$ are defined as follows:

$$D_{mn} = [\int_0^\infty q_3(\hat{D})\hat{D}^m d\hat{D}]/[\int_0^\infty q_3(\hat{D})\hat{D}^n d\hat{D}], \ (m,n) = (3,2) \text{ and } (4,3) \tag{2}$$

If the particles were monodisperse, i.e. $q_3(D) \sim \delta(D-\bar{D})$, we should have $D_{32} = D_{43} = \bar{D}$ ($\delta$ is the Dirac delta function). The fact that $D_{32} \neq D_{43}$ (Fig. 1b) indicates that the protein aggregates are polydisperse – their distribution is close to the lognormal one.

The specific surface area of the SiO$_2$ particles, $A_S$ in Table 1, is given by the expression:

$$A_S = \frac{A}{M} = \frac{\sum_i 4\pi a_i^2}{\sum_i (4/3)\pi a_i^3 \rho_p} = \frac{3}{a_{32}\rho_p}; \ a_{32} \equiv \frac{\sum_i a_i^3}{\sum_i a_i^2} \tag{3}$$

where $A$ and $M$ are the total surface area and the total mass of all particles; $a_i$ is the radius of the $i$-th particle, and $\rho_p$ is the particle mass density; $\rho_p = 2.196$ g/cm$^3$ for amorphous SiO$_2$; $a_{32}$ is the mean particle radius after Sauter [29,30]. The values of $a_{32}$ in Table 1 were calculated from $A_S$ (given by the producer) by using Eq. (2).

The drop radius in particle stabilized (Pickering) emulsions is determined by the phenomenon *limited coalescence* [31-34]: If the surface coverage of the small emulsion drops produced by the homogenizer is not high enough to stabilize them, the drops begin to coalesce. At that, the surface coverage with particles increases, until eventually the coalescence stops. Assuming that practically all available particles are adsorbed on the drop surfaces, one can derive the following formula for the Sauter mean radius, $R_{32}$, of the stable emulsion drops at the end of the process of limited coalescence [34-36]:

$$R_{32} = \frac{4a_{32}\varphi_a\rho_{ps}}{\varphi_p}\frac{\Phi_d}{1-\Phi_d} \tag{4}$$

For the derivation of Eq. (4), see e.g. Appendix A in Ref. [36]; see also Refs. [33,37-39]. As before, $a_{32}$ is the radius of the SiO$_2$ particles; $\varphi_a$ is the area fraction of drop surface that is covered by adsorbed particles; at hexagonal surface close packing of monodisperse spheres, $\varphi_a = \pi/(12)^{1/2} \approx 0.9069$; $\varphi_p$ is the mass fraction of particles in the dispersion; $\rho_{ps} = \rho_p/\rho_s$ is the ratio of the mass densities of particles and solvent (water), and $\Phi_d$ is the volume fraction of the oil drops. Eq. (4) will be used to analyze the data for the size of the produced capsules.

## 4. Experimental results on encapsulation

### 4.1. Optimization of the composition

As mentioned above, in Ref. [1] microcapsules were formed with SiO$_2$ particles of average diameter 3.5 μm. The optimal composition was 5 wt% silica particles + 3 mM KOleate + 0.04 wt% Carbopol 971P + 500 mM NaCl + 10 vol% oil.

In the present study, we are working with nanoparticles, which are much smaller than the protein aggregates – compare the $a_{32}$ values in Table 1 with the aggregate sizes in Fig. 1b. These structural changes lead to different optimal composition for stable capsules. To find this optimal composition, we varied the concentrations of (i) silica particles, (ii) KOleate and (iii) Carbopol 971P. In all experiments, here we worked at fixed concentrations of salt and oil, viz. 500 mM NaCl and 10 vol% oil, which are the same as in Ref. [1].

*Effect of SiO$_2$ particle concentration*. This effect is illustrated in Fig. 3, where the concentration of Ludox TM-50 is varied at fixed concentrations of the other ingredients; the oil phase is limonene. All objects seen in Fig. 3 are emulsion drops; the silica particles are invisible by optical microscopy ($a_{32}$ = 10.5 nm; see Table 1). One sees that with 20 times increase of particle concentration, $\varphi_p$, from 0.05 to 1 wt%, the mean drop size decreases significantly in qualitative agreement with Eq. (4); quantitative comparison is presented in Section 5. This result means that the drop size follows the law of limited coalescence, i.e. we are dealing with particle stabilized emulsions. In other words, despite their nanometer size, and the presence of surfactant and polymer, the silica particles play a central role in the emulsion stabilization, as observed in the numerous studies on Pickering emulsions [13-16,40-44].

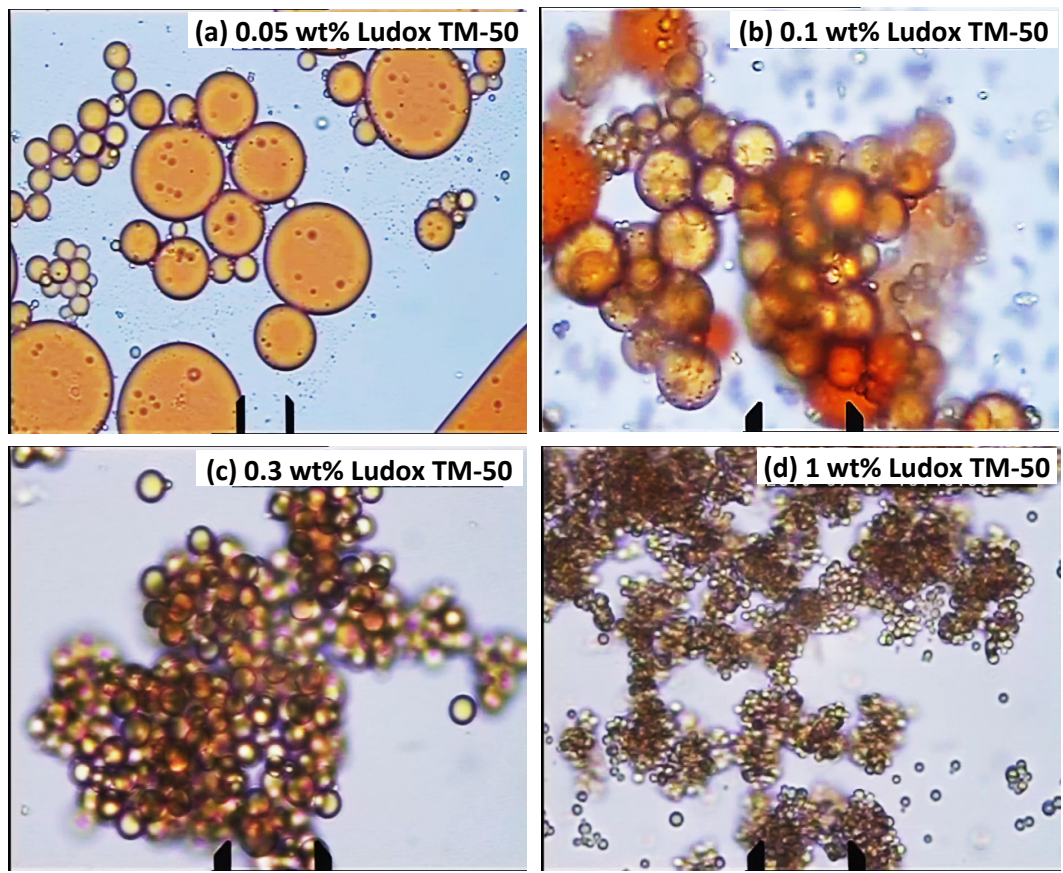

**Fig. 3**. Effect of nanoparticle concentration on the size of the capsules produced from a suspension containing 3 mM KOleate + 0.04 wt% Carbopol 971P + 500 mM NaCl + 10 vol% limonene at several concentrations of Ludox TM-50 particles: (a) 0.05 wt%; (b) 0.1 wt%; (c) 0.3 wt%, and (d) 1 wt%. The micrographs are taken before rinsing with water; the scaling caliper shows 20 μm.

The tendency for decrease of drop size with the rise of particle concentration is observed also with the other kinds of silica particles in Table 1. An additional example is given for Aerosil 300 – see SI Appendix, Fig. S1 (SI = Supplementary Information).

In Fig. 3, most of the silica particles form clusters; see also SI Appendix, Fig. S2. After rinsing with pure water (as described in Section 2.2), these clusters are destroyed, and one obtains separate capsules – oil drops covered with shells of particles, surfactant and polymer; see Section 4.3. In other words, the rinsing removes the excess surfactant and polymer, which bind the capsules into clusters.

*Effect of surfactant – KOleate*. In Ref. [1], with μm-sized silica particles it was shown that KOleate provides reversible hydrophobization of the hydrophilic $SiO_2$ particles. The results presented Figs. 4a–d imply that the same happens also in the case of silica nanoparticles. In this experiment, the concentration of KOleate is varied at fixed concentrations of all other ingredients. The micrographs show the structure of the separated cream.

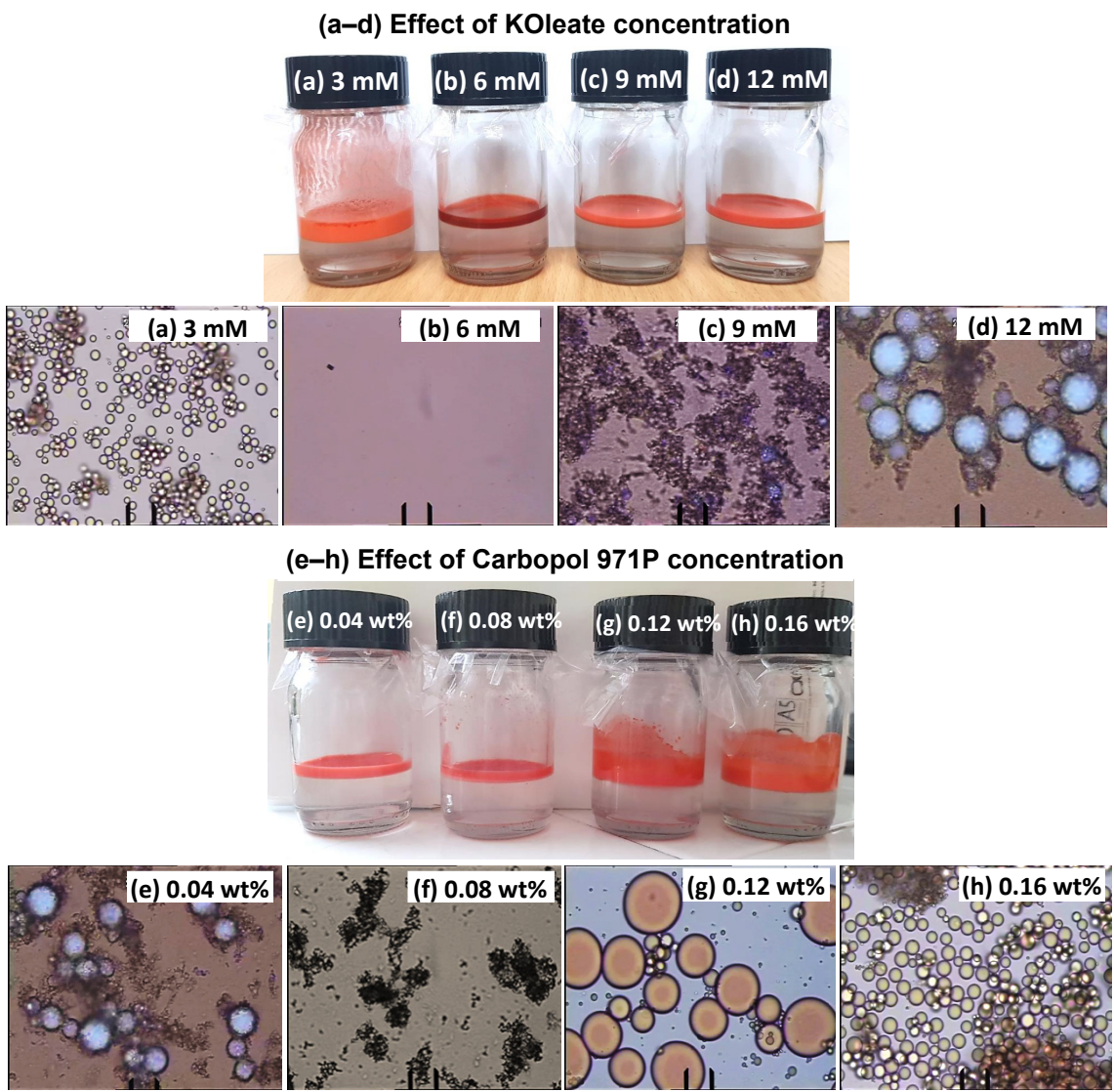

**Fig. 4.** Optimization of the composition – photographs of vials and micrographs of the separated oily phase (colored red with Sudan III). All systems contain 0.1 wt% Ludox SM; 500 mM NaCl, and 10 vol% limonene, whereas the concentrations of KOleate and Carbopol 971P are varied. (a–d) Effect of KOleate concentration (denoted in the pictures) at a fixed concentration of 0.04 wt% Carbopol 971P. (e–h) Effect of Carbopol 971P concentration (denoted in the pictures) at a fixed concentration of 12 mM KOleate. The scaling caliper shows 20 μm.

At the lowest concentration, 3 mM KOleate (Fig. 4a), one observes oil droplets (capsules) dispersed in water. At 6 mM KOleate (Fig. 4b), the cream represents a uniform oil phase, which may contain hydrophobized silica nanoparticles that are too small to be seen by optical microscopy. Figs. 4c and d show that at 9 and 12 mM KOleate the particle hydrophobicity and concentration in the oily phase have become sufficiently high to stabilize the inverse, water-in-oil Pickering emulsions.

The following physical picture emerges from the reported results. The KOleate molecules bind with their headgroups to the surface of the silica particles, which are hydrophobized in this way. If oil drops were missing, the further rise of KOleate concentration would lead to formation of a second surfactant adsorption layer on the silica particles, which would render them hydrophilic again. However, in the presence of oil drops the rise of KOleate concentration, and the related increase of particle hydrophobicity, lead to a transfer of the $SiO_2$ particles from the aqueous into the oily phase. This transfer happens at KOleate concentrations lower than those needed for the formation of second adsorption layer. For this reason, back hydrophilization of the particles with the rise of KOleate concentration is not observed. Note that the degree of hydrophobicity of particle surface (the three-phase contact angle) determines whether direct or inverse emulsion will be formed [12-16,45].

The conclusion is that the concentration of KOleate should be high enough to hydrophobize the silica particle to such an extent that they can stabilize direct Pickering emulsions (Fig. 4a). However, the KOleate concentration should not be too high to transfer the particles in the oily phase, where they could stabilize the inverse water-in-oil emulsions (Figs. 4c and d). Fig. 4b corresponds to an intermediate concentration of KOleate, at which the particles can stabilize neither the direct, nor the inverse Pickering emulsion. These conclusions are supported by additional experimental data shown in SI Appendix, Figs. S3 – S6.

*Effect of polymer – Carbopol 971P*. In Figs. 4e–h, the concentration of Carbopol 971P is varied at fixed concentrations of all other ingredients. The micrographs, which visualize the structure of the separated cream, show a tendency that is the opposite of that in Figs. 4a–d. In Figs. 4e and f, at the lower polymer concentrations the formation of inverse, water-in-oil emulsions is seen. Fig. 4g shows a direct emulsion with relatively large drops. In Fig. 4h, at the highest used concentration of 0.16 wt% polymer, a finer emulsion of lower polydispersity is seen.

The results for Ludox SM in Figs. 4e–h indicate the presence of well pronounced surfactant-polymer interaction. At the lowest polymer concentration, there is an excess of surfactant, which strongly hydrophobizes the particles, so that they enter the oily phase and stabilize the inverse Pickering emulsion (Figs. 4e and f). With the rise of polymer concentration, a greater fraction of surfactant molecules bind to the polymer, so that the surfactant adsorption on the particles (and their hydrophobicity) becomes lower. In other words, there is a "competition" between the polymer and particles with respect to binding of surfactant molecules. For this reason, the rise of polymer concentration is accompanied with a decrease of the amount of surfactant bound to the silica particles, and the changes in Figs. 4e–h are in direction opposite to those in Figs. 4a–d. Similar results for Ludox HS-30, Ludox TM-50 and Aerosil 300 at various polymer concentrations are shown in SI Appendix, Figs. S7 – S9.

In particular, the relatively large drops in Fig. 4g can be explained with the fact that the particles are still too hydrophobic to stabilize well a direct Pickering emulsion [13-16]. In contrast, in Fig. 4h the ratio of the surfactant and polymer concentrations corresponds to an appropriate degree of particle hydrophobicity, so that fine and relatively monodisperse droplets are formed. As *optimal composition* for the formation of oil-in-water microcapsules, we selected that corresponding to Fig. 4h, viz. 0.1 wt% $SiO_2$ nanoparticles; 12 mM KOleate; 0.16 wt% Carbopol 971P; 500 mM NaCl, and 10 vol% oil. This composition is used in our subsequent experiments. Upon rinsing the microcapsules with pure water, the particle clusters (some of them are seen in Fig. 4h) are disassembled and one obtains separate stable and relatively monodispersed capsules in the aqueous phase; see Section 4.3.

*4.2. Discussion*

One of the conclusions from Fig. 4 is that the polymer binds a part of the surfactant molecules available in the solution and, thus, controls the amount of surfactant bound to the silica particles and their hydrophobicity. If this is the only role of polymer, then one could decrease the concentration of surfactant and work without any polymer in the solution.

To investigate this possibility, we carried out experiments at surfactant concentrations lower than 12 mM (viz. 3, 6 and 9 mM KOleate + 0.1 wt% Ludox SM) without any polymer. The results (SI Appendix, Fig. S10) show that at 3 mM KOleate emulsion does not form at all – the oil is separated as a layer above the aqueous phase. At 6 and 9 mM KOleate, Pickering emulsion is formed, but after rinsing with water it is destroyed – layer of oil is released above the aqueous phase. The latter represents a common emulsion stabilized by residual surfactant molecules (a single rinsing with water cannot remove the whole amount of surfactant in the dispersion). Thus, we arrive to the conclusion that the presence of polymer (Carbopol 971P) is a necessary condition for the formation of stable capsules, which can survive rinsing with water.

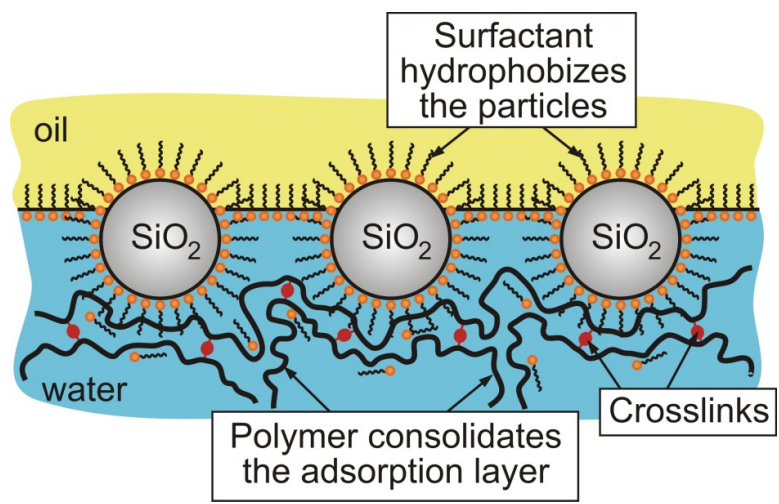

**Fig. 5.** Sketch of the capsule shell: Silica nanoparticles, hydrophobized by adsorbed surfactant (KOleate) molecules, are attached to the oil/water interface. Segments of the polymer molecules (aggregates) are attached to the hydrophobized particle/water interface. The thick layer of adsorbed cross-linked polymer consolidates the composite adsorption layer and blocks the desorption of surfactant upon rinsing with water.

The physical picture emerging from these experiments is illustrated in Fig. 5. The silica particles are hydrophobized by adsorbed surfactant (KOleate), which ensures their attachment to the oil/water interface. As already mentioned, the surfactant concentration is insufficient for the formation of second adsorption layer on the particles; before that they would enter the oily phase, as experimentally observed (Fig. 4). Because the particle surfaces facing the aqueous phase are hydrophobized, they become adsorption cites for segments of the polymer molecules. The reason for such adsorption is the hydrophobic interaction between the tails of the surfactant molecules bound to the particles and the hydrophobic backbone of the polymer chain (Fig. 5). In this way, the particle adsorption layer is consolidated by bound polymer molecules, which are much bigger than the nanoparticles (Fig. 1b). To some extent, such shell morphology resembles the cell wall of the gram-positive bacteria: the particle adsorption layer is a counterpart of the cytoplasmic membrane, whereas the outer negatively charged layer of crosslinked polymer resembles the peptidoglycan – see e.g. Ref. [46], Fig. 5 therein.

The formed complex of nanoparticles + surfactant + polymer turns out to be stable against rinsing with pure water. In other words, the layer of bound polymer retains the surfactant molecules in the adsorption shell and blocks their migration toward the bulk aqueous phase.

At higher concentrations, the polymer enhances the capsule flocculation, i.e. the formation of clusters from capsules (SI Appendix, Fig. S11). At fixed nanoparticle and surfactant concentrations, the concentration of polymer affects the size of the produced capsules (SI Appendix, Fig. S12). At too high polymer concentration, the capsules become polydisperse (SI Appendix, Figs, S13 and S14). This could be due to binding of a considerable part of the surfactant to the polymer aggregates, which lowers the surfactant adsorption on the nanoparticles; decreases their hydrophobicity and the density of their adsorption layer on the oil/water interface, which in a final reckoning decreases the stability of the formed capsules. Stable and relatively monodisperse capsules are formed in the vicinity of the optimal composition, which is 0.1 wt% $SiO_2$ nanoparticles; 12 mM KOleate; 0.16 wt% Carbopol 971P; 500 mM NaCl, and 10 vol% oil; see above.

The difference between the particle size in Ref. [1] and here, viz. μm-sized particles vs. nanoparticles, leads also to differences in the structure of the formed composite shells. In Ref. [1], the particles were bigger than the polymer aggregates, which fill the gaps between the adsorbed particles ("brick-and-mortar" structure) and, thus, prevent the surfactant desorption and leakage of oil. In the present study, the protein aggregates are much bigger than the silica nanoparticles, and their adsorption layer plays the role of consolidating network that additionally stabilizes the encapsulating shell. This structural alteration leads to functional differences between the capsules produced with micro- and nanoparticles. In the latter case, the capsules are smaller in size and more stable against the rise of pH and the treatment with detergent solution (Section 6). In particular, the capsule destabilization at pH > 6 for microparticles and pH > 11 for nanoparticles can be explained with the desorption of the polymer in the former case and of the nanoparticles in the latter case. Moreover, the different structures lead to a different optimal composition of particles, surfactant and polymer – see above.

*4.3. Encapsulation of different fragrances and oils*

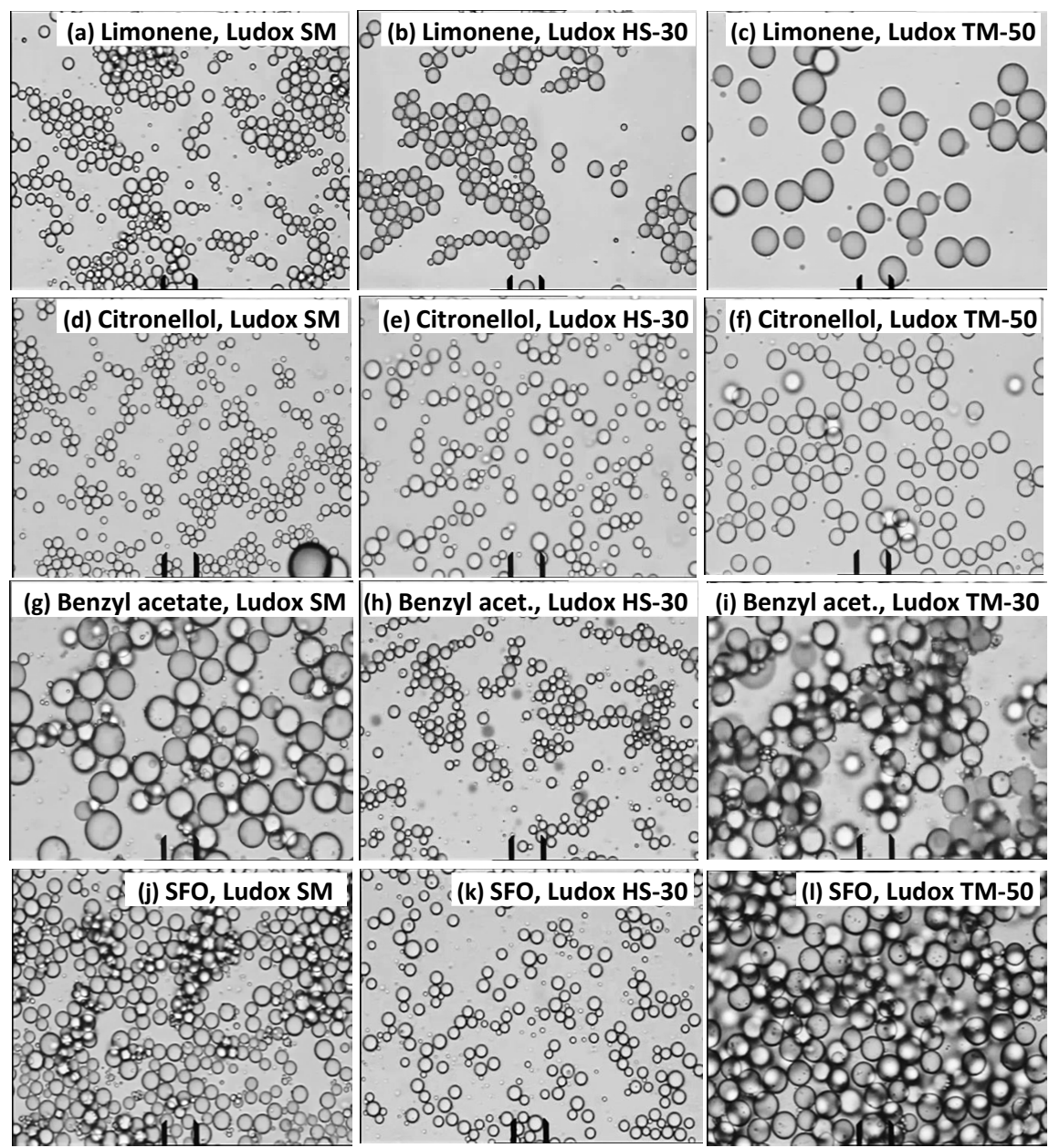

**Fig. 6.** Micrographs of the produced capsules after rinsing. The composition is 0.1 wt% silica particles + 12 mM KOleate + 0.16 wt% Carbopol 971P + 500 mM NaCl + 10 vol% oil. The kinds of oil and silica particles are specified on the micrographs. The scaling caliper shows 20 μm.

Fig. 6 shows micrographs of the stable capsules (after rinsing with water) produced at the optimal composition with three types of particles (Ludox SM, Ludox HS-30 and Ludox TM-50) and with four types of oils (limonene, citronellol, benzyl acetate and SFO). Qualitatively, the capsule size follows the predictions of Eq. (4) that (i) the capsule radius $R_{32}$ should be independent of the type of oil and (ii) that capsule size should grow proportional to the particle size, $R_{32} \propto a_{32}$. Two exclusions are the capsules of benzyl acetate and SFO with Ludox SM, which are bigger than expected. A possible explanation could be that in these systems the Ludox SM particles are slightly aggregated (i.e. behave as particles with larger effective radius $a_{32}$), which would lead also to larger $R_{32}$. Insofar as we are dealing with particle stabilized emulsions, we could expect that Eq. (4) determines the main trend of the experimental data. However, deviations from Eq. (4) are also possible, because it does not take into account effects as particle aggregation and adsorption of segments of the polymer between the particles. More detailed discussion can be found in Section 5 on the basis of the analysis of data for $R_{32}$.

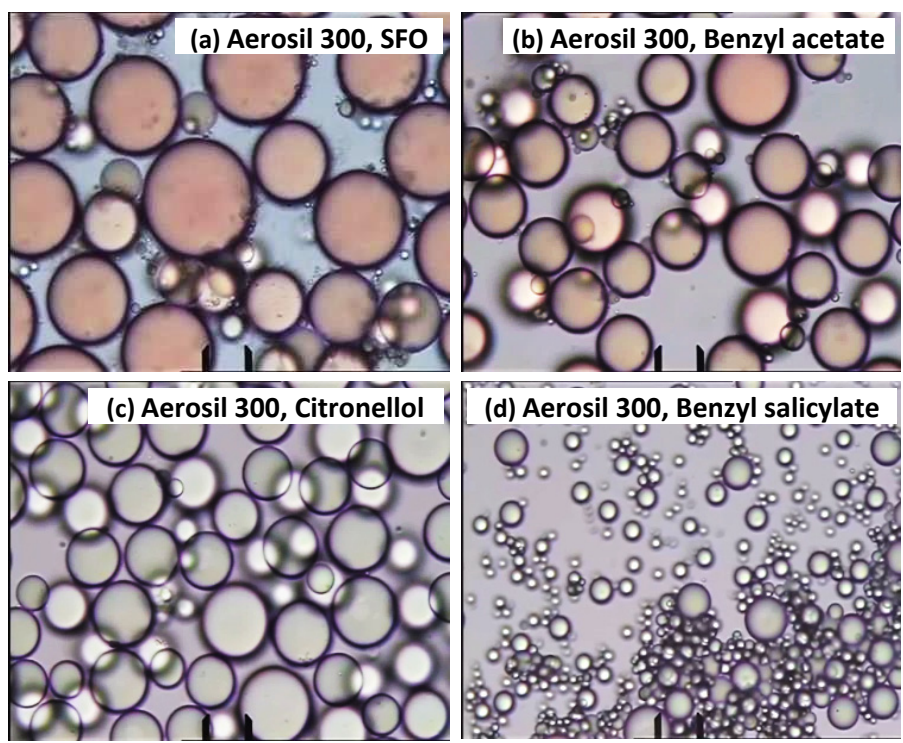

**Fig. 7.** Micrographs of capsules with Aerosil 300 after rinsing. The composition is 0.1 wt% Aerosil 300 + 12 mM KOleate + 0.16 wt% Carbopol 971P + 500 mM NaCl + 10 vol% oil. The kind of oil is specified on the micrographs. The scaling caliper shows 20 μm.

Fig. 7 shows that stable capsules can be formed also with the Aerosil 300, despite the fact that in this case the particles represent polydisperse fractal-like aggregates (of average size ca. 180 nm) from adherent primary $SiO_2$ spheres. As seen in this figure, with Aerosil 300 one can encapsulate SFO, citronellol, benzyl acetate and benzyl salicylate, as well as limonene (SI Appendix, Fig. S1). In particular, in Figs. 7a and b one sees that the surface of the capsules is hairy, which could be explained with the presence of a dense outermost adsorption layer of µm-sized polymer aggregates (see Fig. 1b). In this case, the thickness of the polymer adsorption layer turns out to be sufficiently large to be observed by optical microscopy and to visualize the model depicted in Fig. 5. We could hypothesize that the formation of such thick polymer layer is promoted by the larger surface area of the fractal-like aggregates of Aerosil 300, which offer more adsorption cites, as well as by the nature of the oil, which is accessible for contact with the polymer in the openings between the adsorbed irregularly shaped aggregates.

As every method, the approach developed in the present study has some limitations. We could not succeed to produce good microcapsules (like those in Figs. 6 and 7) filled with liquid paraffins like tetradecane and hexadecane – alkanes of chainlength close to that of KOleate. A possible explanation could be that the oleate is well soluble in these organic liquids that leads to its desorption from the particle/oil interface (see Fig. 5). As a result, the particles recover their hydrophilicity and detach from the oil/water interface into the aqueous phase. In this way, the composite encapsulating shell is destroyed. In principle, this problem could be overcome by replacing the KOleate with another carboxylate of appropriate hydrocarbon chain; this could be a subject of subsequent study.

## 5. Quantitative analysis of capsule size

### 5.1. Effects of particle size, surfactant and polymer

With parameter values corresponding to the optimal composition, $\varphi_p = 0.001$, $\Phi_d = 0.1$ and $\rho_{ps} = 2.196$, and for densely packed particle adsorption layer, $\varphi_a \approx 0.9069$, Eq. (4) acquires the form:

$$R_{32} \approx 885.13\, a_{32} \tag{5}$$

The coefficient in Eq. (5) is dimensionless. According to this equation, nanoparticles of larger mean radius $a_{32}$ stabilize emulsion drops (capsules) of larger mean radius $R_{32}$.

In Fig. 8a, the prediction of Eq. (5) (the solid straight line) is compared with data for $R_{32}$ for capsules from limonene and linalool stabilized with the three kinds of Ludox particles of mean radii $a_{32}$ given in Table 1. These are the capsules illustrated with micrographs in the first two rows of Fig. 6.

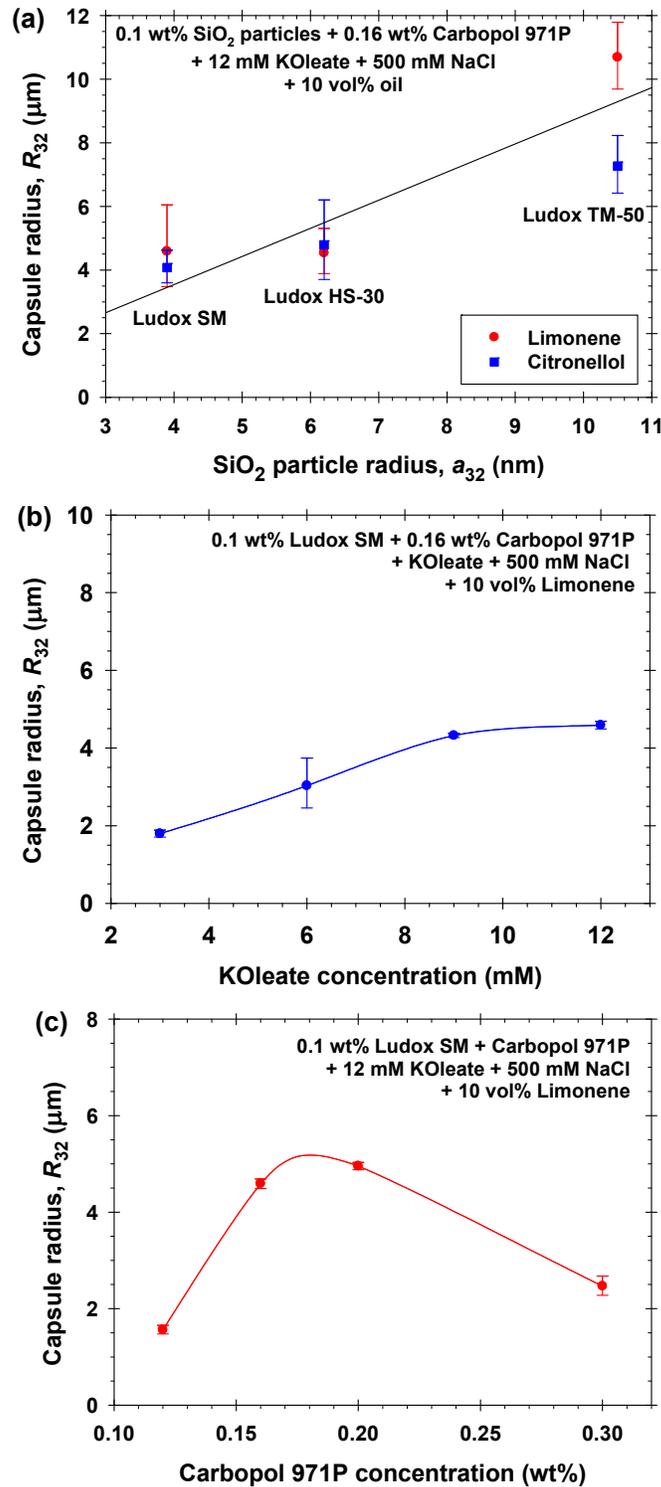

**Fig. 8.** Plots of experimental data for the mean capsule radius: (a) $R_{32}$ vs. the mean radius of the silica nanoparticles, $a_{32}$; the solid line is drawn in accordance with Eq. (5); (b) $R_{32}$ vs. the KOleate concentration, (c) $R_{32}$ vs. the concentration of polymer – Carbopol 971P. The ingredient concentrations are denoted on the graphs. In Figs. 8b and c, the lines are guides to the eye.

For each system, the diameters of 500 capsules were measured and, next, their mean radius $R_{32}$ was calculated – see the experimental points in Fig. 8a. One sees that as order of magnitude and trend, the experimental data follow the prediction of Eq. (5). The error bars in Fig. 8a present the standard error of the mean, which is not so large insofar as the capsules are relatively monodisperse. Deviations from Eq. (5) are present because the experimental system contains also surfactant and polymer, which is not taken into account in Eq. (5). Indeed, Figs. 8b and c demonstrate that the capsule size depends also on the concentrations of K-Oleate and Carbopol 971P.

*5.2. Discussion on the factors determining capsule size*

*Negative* deviations from Eq. (5) (from the straight line in Fig. 8a) could be explained with the fact that the adsorption of fragments from the polymer on the oil/water interface leads to lower surface coverage, $\varphi_a$, by nanoparticles, which in turns, leads to lower $R_{32}$ values; see Eq. (4). The influence of this effect is expected to become stronger with the rise of polymer concentration. In contrast, with the rise of surfactant concentration the polymer becomes increasingly more hydrophilic (due to bound surfactant molecules), which suppresses the binding of polymer fragments to the oil-water interface.

*Positive* deviations from Eq. (5) can be due to the action of other two factors, or to their interplay. (i) Aggregation of the nanoparticles leads to greater effective particle radius $a_{32}$, and therefrom – to greater $R_{32}$, at the same particle volume fraction, $\varphi_p$. (ii) Binding of nanoparticles to the protein aggregates in the bulk (mediated by surfactant molecules) leads to lower effective volume fraction, $\varphi_p$, of the "free" particles and, consequently – to greater $R_{32}$; see Eq. (4).

Fig. 8b shows that $R_{32}$ first increases and then levels off with the rise of surfactant concentration. This behavior can be explained with the gradual hydrophilization of protein molecules with the rise of surfactant concentration, which suppresses the adsorption of their segments to the oil/water interface between the particles; see the paragraph for negative deviations from Eq. (5). Partial aggregation of the hydrophobized nanoparticles (engendered by the hydrophobic interactions between them) could also lead to higher $R_{32}$ values. The adsorption of particle aggregates is equivalent to formation of particle layer, which is thicker than monolayer.

In Fig. 8c, the non-monotonic dependence of $R_{32}$ on polymer concentration can be explained with the competition of two effects acting in opposite directions. At lower polymer concentrations, the effect of aggregation of the hydrophobized nanoparticles (leading to increase of the effective $a_{32}$) prevails and $R_{32}$ increases. However, at higher polymer concentrations, this effect is suppressed because the polymer binds a part of surfactant that causes particle hydrophobization. In addition, with the rise of polymer concentration the adsorption of polymer segments between the particles at the oil/water interface [which causes negative deviations from Eq. (5)] becomes predominant; this effect leads to lower $\varphi_a$ and smaller $R_{32}$; see Eq. (4).

## 6. Capsule destruction and cargo release

### 6.1. Effect of pH on microcapsule stability

One could expect that at sufficiently high pH the ingredients in the encapsulating shell will acquire negative electric charge, which will lead to shell destruction and cargo release. In other words, the produced capsules are expected to be pH-responsive. To verify this, we carried out the following experiment with all kinds of nanoparticles in Table 1. After the gravitational separation of the capsules, the adjacent water phase was replaced with water of different pH values ranging from 3 to 12.5, which have been adjusted with HCl or NaOH. Next, the vial was shaken by hand. As seen in Fig. 9a, the capsules remain stable in a wide range, $3 \leq \text{pH} \leq 10$. Indications for release of the encapsulated material (cargo release) are seen for pH $\geq$ 11, where the aqueous phase is turbid and has reddish color because of the formation of a common (surfactant stabilized) emulsion from the released oil, which contains the dye Sudan III. The microscopic observations of samples from the cream show stable capsules at $3 \leq \text{pH} \leq 10$, and a polydisperse mixture of much bigger and smaller emulsion droplets produced after the capsule destruction at pH $\geq$ 11; see SI Appendix, Figs. S15 – S18.

### 6.2. Discussion on the pH effect

It should be noted that in our previous study with a similar system [1], the $SiO_2$ microparticles (of diameter = 3.5 μm), and the cargo release was observed for pH > 6 at which Carbopol 971P acquires significant negative charge. Correspondingly, the cargo release was interpreted as a consequence of polymer charging and desorption from the spaces between the microparticles, which triggers destruction of the encapsulating shell.

In the present experiments with *nanoparticles*, the produced capsules turn out to be more resistant against the rise of pH, insofar as the cargo release occurs at pH $\geq$ 11 (Fig. 9a).

Both Carbopol and KOleate in the bulk acquire negative charge (deprotonation of carboxylic groups) at significantly lower pH values; for the medium and long-chain fatty acids, $pK_a \approx 4.7$ [47,48]. Hence, the only possible explanation for the transition from stable to unstable capsules is that at pH ≥ 11 the $SiO_2$ nanoparticles acquire so high negative surface charge that it becomes energetically favorable they to detach from the oil/water interface and to go to the aqueous phase. This explanation is consistent with the experimentally detected variation of the zeta-potential at high pH – see Fig. 1a. The difference between the values of the threshold pH for capsule destruction for microparticles [1] and nanoparticles (6 vs. 11), is most probably related to the different relative sizes (and functions) of the polymer aggregates and $SiO_2$ particles.

**(a) Effect of pH**

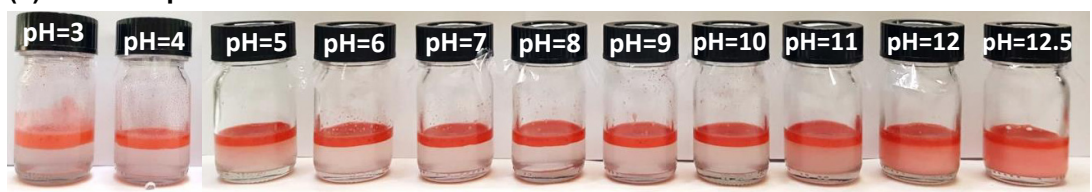

**(b) Effect of the addition of anionic detergent, SLES-1EO**

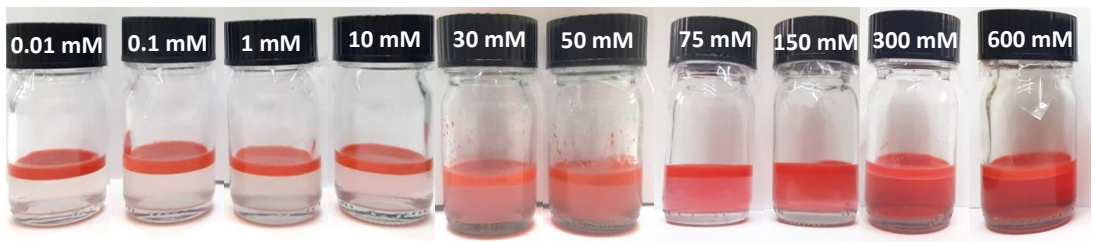

**Fig. 9.** Photographs of vials with microcapsules: (a) After rinsing with water of different pH values denoted on the caps of the vials; the particles are Ludox HS-30. (b) After rinsing with water containing different concentrations of SLES-1EO denoted on the caps of the vials; the particles are Ludox SM. The used composition is 0.1 wt% silica particles + 12 mM KOleate + 0.16 wt% Carbopol 971P + 500 mM NaCl + 10 vol% SFO. The microcapsules are in the cream. Subphase of red color indicates capsule destruction.

*6.3. Effect of anionic detergent on microcapsule stability*

Fig. 9b illustrates the effect of rinsing of the capsules with solutions containing different concentrations of the anionic detergent, SLES-1EO. One sees that the transition between stable and unstable capsules occurs between 10 and 30 mM, i.e. at ca. 20 mM SLES-1EO. For SLES-1EO concentration ≥ 30 mM, the aqueous sub-phase becomes turbid and acquires red color because of the capsule destruction, cargo release and formation of a common surfactant stabilized emulsion.

SLES-1EO adsorbs on the polymer aggregates, hydrophilizes them and brings them in the bulk water phase. This is sufficient to destroy the capsules (see Section 4.2). In addition, SLES-1EO is able to hydrophilize the KOleate-coated $SiO_2$ nanoparticles (Fig. 5) and to detach them from the oil-water interface.

For the *microparticle* stabilized capsules in Ref. [1], the stability/instability transitional concentration of SLES-1EO was 0.1 mM, whereas for the *nanoparticle* stabilized capsules investigated here this concentration is 20 mM, i.e. 200 times higher. Hence, the use of nanoparticles leads to higher stability of the produced capsules against rinsing with anionic detergent solutions.

## 7. Conclusions

In our previous study [1], we report that oils and fragrances can be encapsulated by using composite shells of μm-sized silica particles, surfactant and polymer. In the present paper, we demonstrate that similar core-shell structures can be obtained with 500 times smaller silica nanoparticles. The properties of the produced capsules are investigated.

The template for creation of the core-shell structure is a particle stabilized (Pickering) emulsion [4,17]. The surfactant, potassium oleate, adsorbs on the nanoparticles and leads to their reversible hydrophobization and adsorption on the oil/water interface. The adsorbed nanoparticles represent the skeleton of the capsule shell. The results indicate that the outer part of the shell represents a layer from the crosslinked polymer Carbopol 971P (see Figs. 5, 7a and b).

The procedure of encapsulation is simple and includes single homogenization by ultrasound of the formulation that contains all ingredients together. In this respect, it is different from the layer-by-layer technique [8-10]. The produced capsules have mean radius in the range $2 < R_{32} < 11$ μm. By order of magnitude and trend, the capsule size follows the law of limited coalescence, Eq. (4), with respect to the dependence on nanoparticle size and concentration (Section 5). The composite structure of the shells leads also to dependence of $R_{32}$ on the concentrations of added polymer and surfactant (Fig. 8).

The produced microcapsules are stable in aqueous environment and survive rinsing with water of pH in the range 3 – 10. However, if dispersed in water of pH ≥ 11, the microcapsules are destabilized and release their cargo, i.e., they are pH responsive. The capsules can be destroyed also by rinsing with a solution of detergent at a sufficiently high concentration (Section 6).

With the optimized formulation, various fragrances and oils, such as limonene, citronellol, benzyl acetate, and sunflower seed oil were encapsulated (Figs. 6 and 7). The developed methodology could find applications in any field, in which reversible encapsulation of oily substances is needed.

**Acknowledgments**

The authors gratefully acknowledge the support from the National Science Fund of Bulgaria, Grant No. DN 09/8/2016 and from the Operational Programme "Science and Education for Smart Growth", Bulgaria, grant number BG05M2OP001-1.002-0012. They thank Assoc. Prof. Dr. Krastanka Marinova for the helpful discussions and Dr. Mihail Georgiev for the particle sizing measurements with Carbopol 971P.

**Appendix. Supplementary Information**

Supplementary data associated with this article can be found in the online version, at http://dx.doi.org/10.1016/j...

# Supplementary Information

for the paper

# Encapsulation of fragrances and oils by core-shell structures from silica nanoparticles, surfactant and polymer: Effect of particle size

Tatiana G. Slavova, Gergana M. Radulova, Peter A. Kralchevsky*, Krassimir D. Danov

*Department of Chemical and Pharmaceutical Engineering, Faculty of Chemistry and Pharmacy, Sofia University, Sofia 1164, Bulgaria*

* Corresponding author; email address: pk@lcpe.uni-sofia.bg

# Appendix

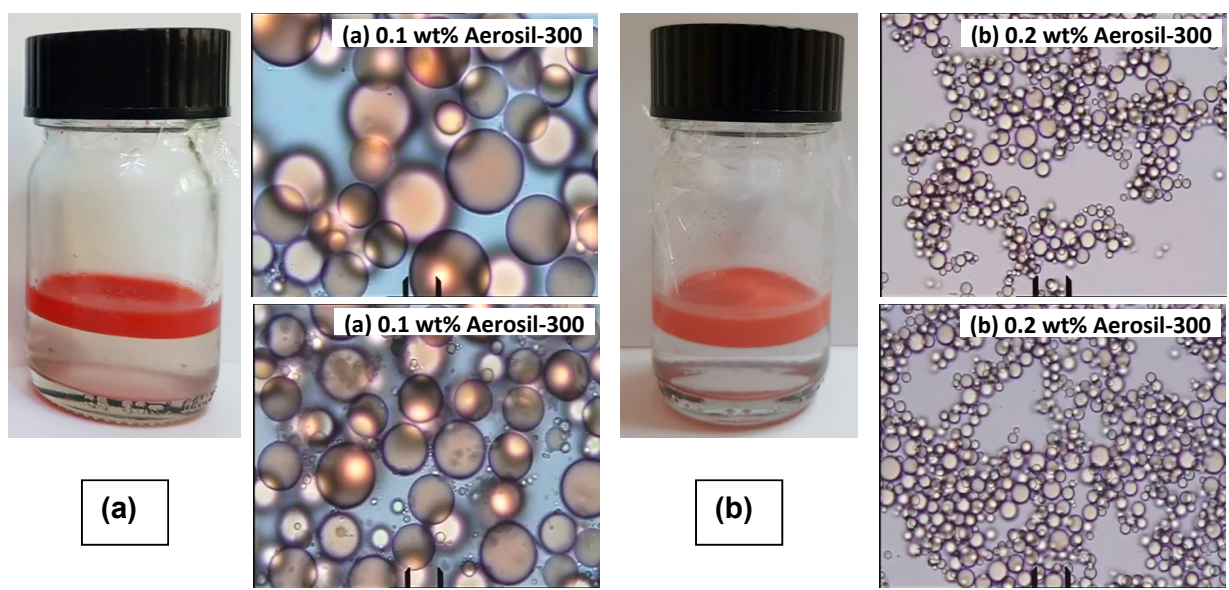

**Fig. S1.** Change in the size of the capsules produced from 10 vol% limonene upon doubling the concentrations of the other ingredients – photographs of the vials and micrographs of the capsules in the cream. (a) Composition: 0.1 wt% Aerosil 300 + 12 mM KOleate + 0.16 wt% Carbopol 971P + 500 mM NaCl + 10 vol% limonene. (b) Composition: 0.2 wt% Aerosil 300 + 24 mM KOleate + 0.32 wt% Carbopol 971P + 1 M NaCl + 10 vol% limonene. In the micrographs, the scaling caliper shows 20 μm.

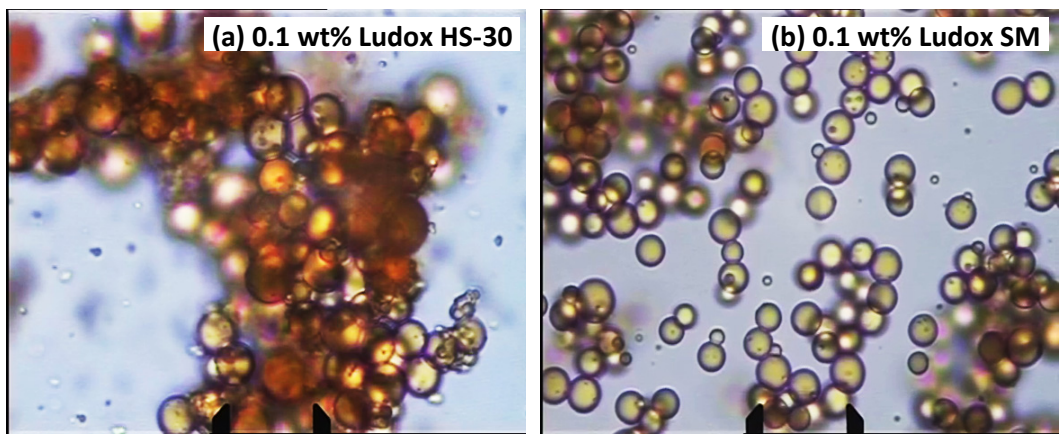

**Fig. S2.** Micrographs of capsules produced from dispersions containing 0.1 wt % silica nanoparticles + 3 mM KOleate + 0.04 wt% Carbopol 971P + 500 mM NaCl + 10 vol% limonene. The nanoparticles are (a) Ludox HS-30 and (b) Ludox SM. The micrographs are taken before rinsing with water; after rinsing, the clusters of capsules decompose to separate capsules. The scaling caliper shows 20 μm.

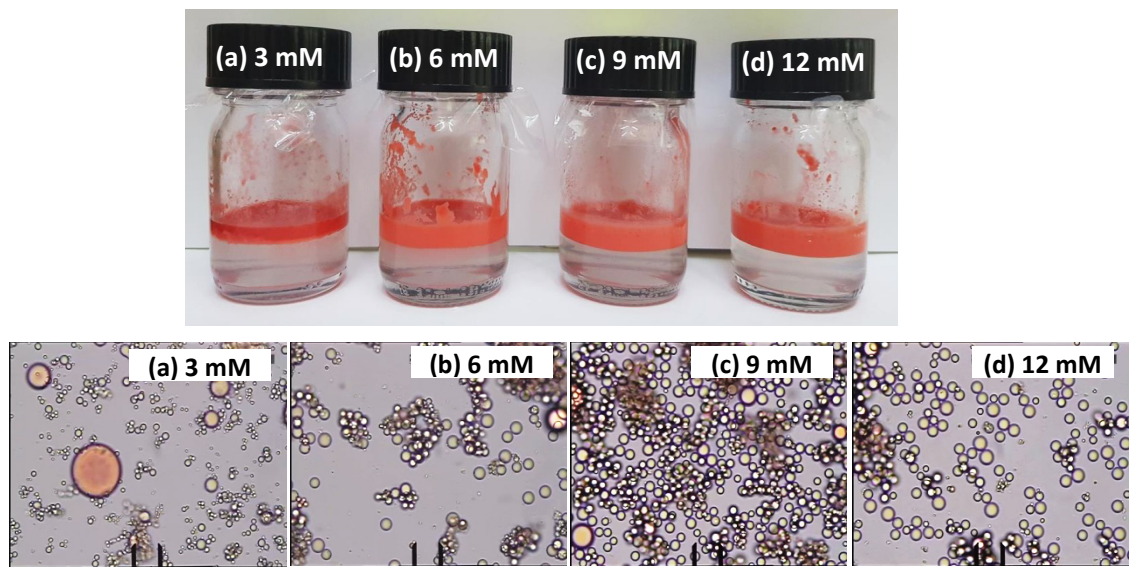

**Fig. S3.** <u>Ludox SM nanoparticles</u>: Effect of KOleate concentration (denoted in the pictures) at a *higher* fixed polymer concentration (as compared to Figs. 4a–d in the main text), viz. 0.16 wt% Carbopol 971P; photographs of the vials and micrographs of the capsules in the cream. The composition is 0.1 wt% Ludox SM + KOleate + 0.16 wt% Carbopol 971P + 500 mM NaCl + 10 vol% limonene; the scaling caliper shows 20 μm. Because of the higher polymer concentration, a greater part of KOleate is bound to the polymer aggregates. As a result, the nanoparticles are not over-hydrophobized and do not enter the oil phase, as in Figs. 4a–d in the main text.

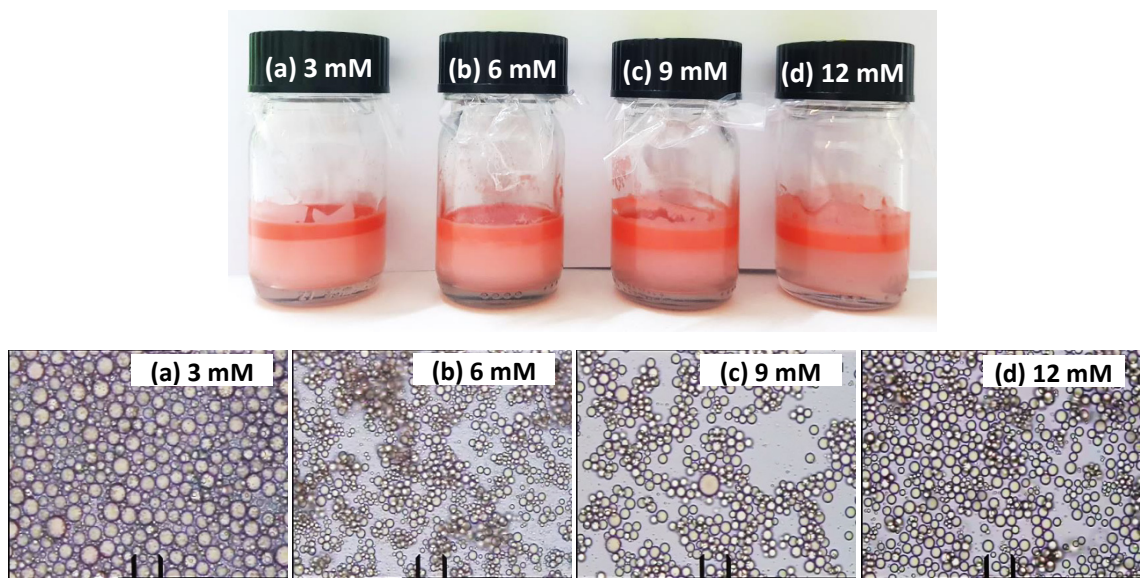

**Fig. S4.** <u>Ludox HS-30 nanoparticles</u>: Effect of KOleate concentration (denoted in the pictures) at a *higher* fixed polymer concentration (as compared to Figs. 4a–d in the main text), viz. 0.16 wt% Carbopol 971P; photographs of the vials and micrographs of the capsules in the cream. The composition is 0.1 wt% Ludox HS-30 + KOleate + 0.16 wt% Carbopol 971P + 500 mM NaCl + 10 vol% limonene. In the micrographs, the scaling caliper shows 20 μm.

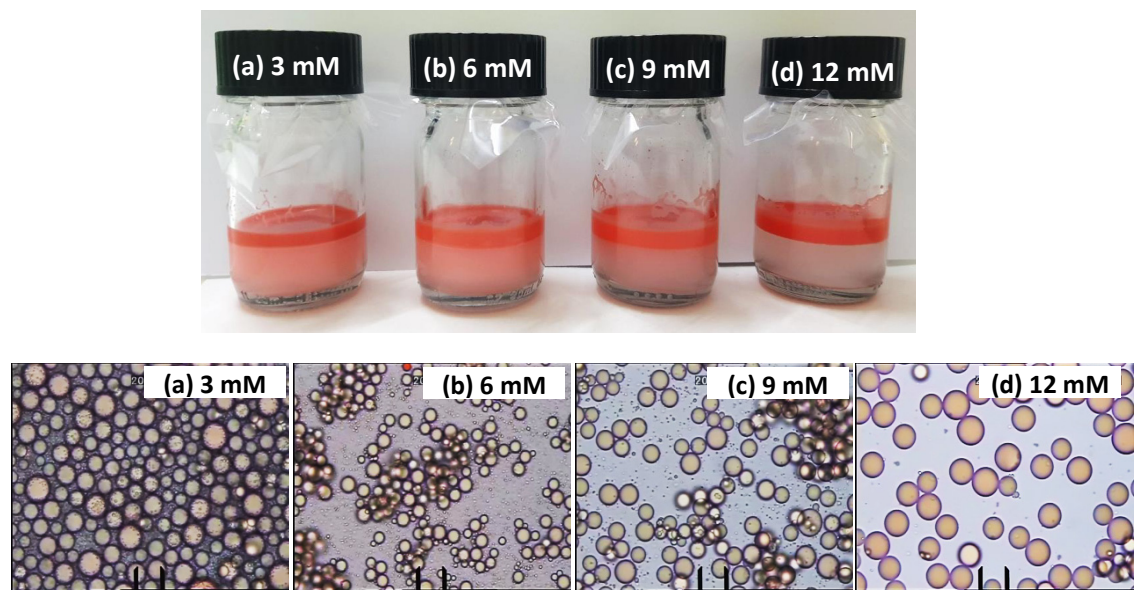

**Fig. S5.** <u>Ludox TM-50 nanoparticles</u>: Effect of KOleate concentration (denoted in the pictures) at a *higher* fixed polymer concentration (as compared to Figs. 4a–d in the main text), viz. 0.16 wt% Carbopol 971P; photographs of the vials and micrographs of the capsules in the cream. The composition is 0.1 wt% Ludox TM-50 + KOleate + 0.16 wt% Carbopol 971P + 500 mM NaCl + 10 vol% limonene; the scaling caliper shows 20 μm.

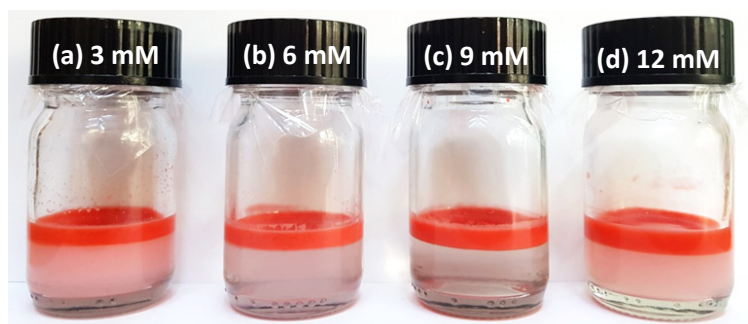

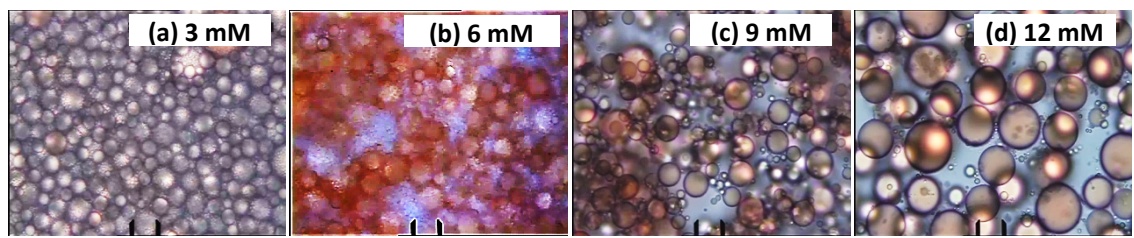

**Fig. S6.** <u>Aerosil 300 nanoparticles</u>: Effect of KOleate concentration (denoted in the pictures) at a *higher* fixed polymer concentration (as compared to Figs. 4a–d in the main text), viz. 0.16 wt% Carbopol 971P; photographs of the vials and micrographs of the capsules in the cream. The composition is 0.1 wt% Aerosil 300 + KOleate + 0.16 wt% Carbopol 971P + 500 mM NaCl + 10 vol% limonene; the scaling caliper shows 20 μm.

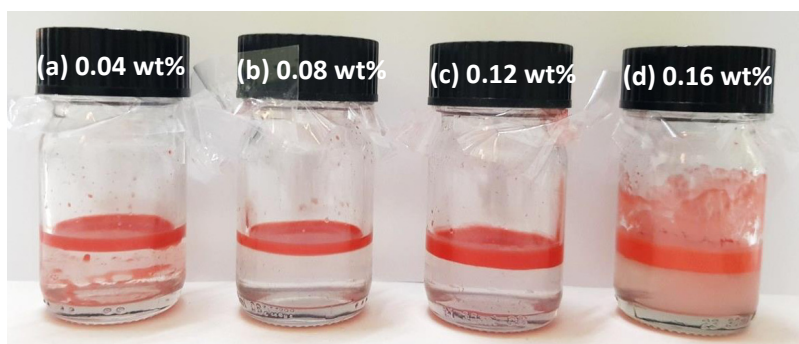

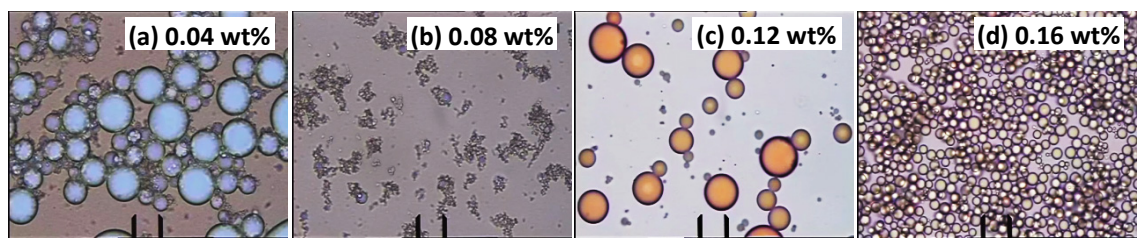

**Fig. S7.** <u>Ludox HS-30 nanoparticles</u>: An experiment analogous to Fig. 4e–h in the main text with other nanoparticles – effect of Carbopol 971P concentration (denoted in the pictures). The other ingredients are 0.1 wt% Ludox HS-30 + 12 mM KOleate + 500 mM NaCl + 10 vol% limonene; photographs show of the vials and micrographs of the capsules in the cream; the scaling caliper shows 20 μm.

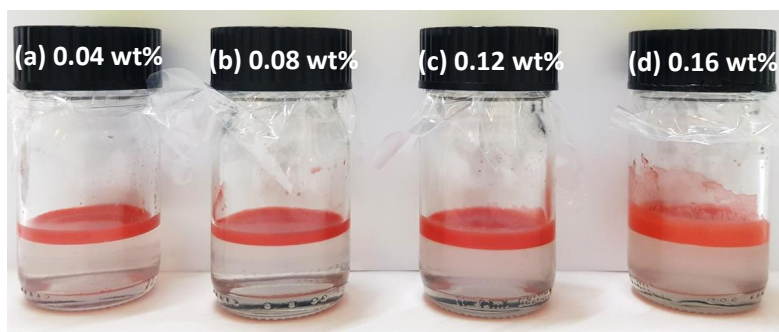
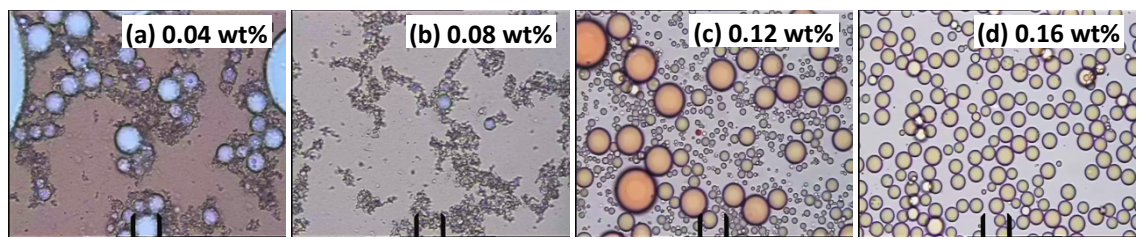

**Fig. S8.** <u>Ludox TM-50 nanoparticles</u>: An experiment analogous to Fig. 4e–h in the main text with other nanoparticles – effect of Carbopol 971P concentration (denoted in the pictures). The other ingredients are 0.1 wt% Ludox TM-50 + 12 mM KOleate + 500 mM NaCl + 10 vol% limonene; photographs show of the vials and micrographs of the capsules in the cream; the scaling caliper shows 20 μm.

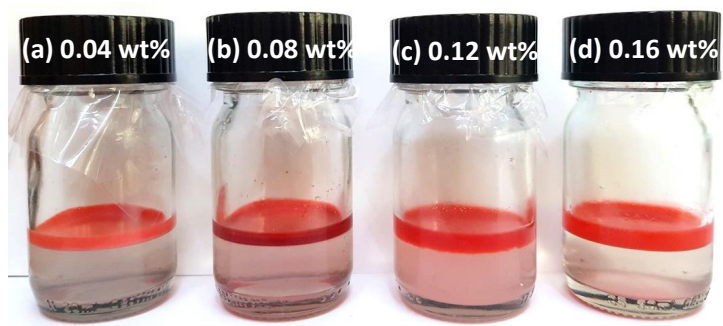
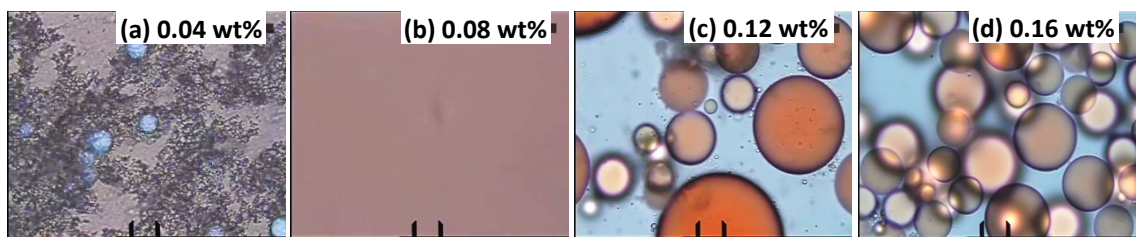

**Fig. S9.** <u>Aerosil 300 nanoparticles</u>: An experiment analogous to Fig. 4e–h in the main text with other nanoparticles – effect of Carbopol 971P concentration (denoted in the pictures). The other ingredients are 0.1 wt% Aerosil 300 + 12 mM KOleate + 500 mM NaCl + 10 vol% limonene; photographs show of the vials and micrographs of the capsules in the cream; the scaling caliper shows 20 μm.

0.1 wt% Ludox SM + 3 / 6 / 9 M KOleate + 500 mM NaCl + 10 vol% limonene

**(a) Before rinsing**

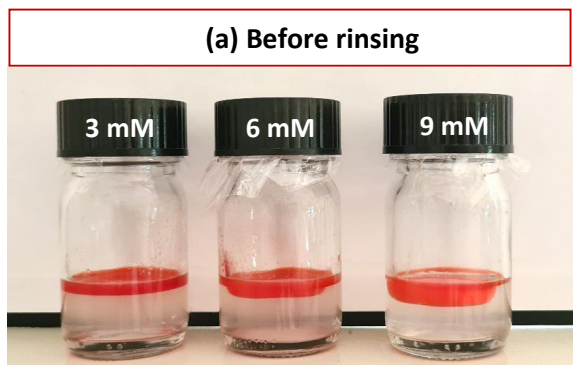

**(b) After rinsing**

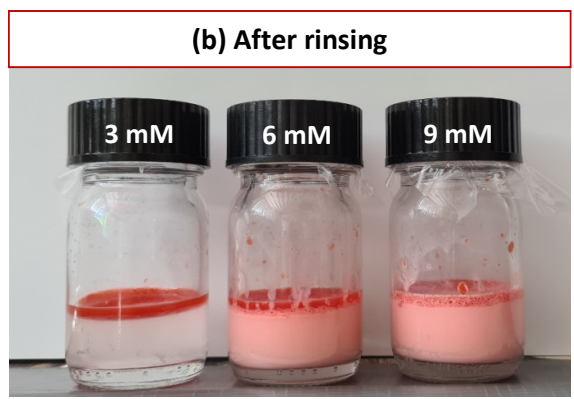

**(c) Micrograps**

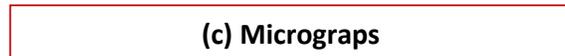

0.1 wt% Ludox SM + 6 mM KOleate + 500 mM NaCl + 10 vol% limonene

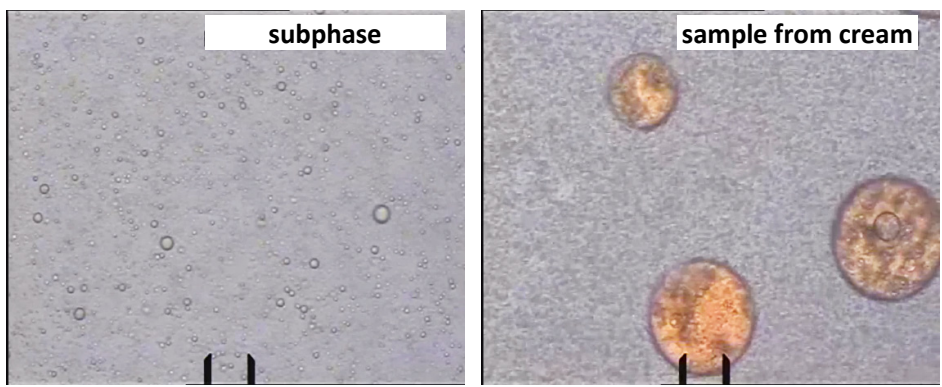

**Fig. S10**. An attempt to produce capsules <u>without polymer Carbopol 971P</u>; the concentration of KOleate (shown on the caps of the vials) is varied. (a) The phase-separated dispersions before the rinsing with water: capsules are formed at 6 and 9 mM KOleate, whereas at 3 mM KOleate capsules are not formed at all – the separated layer is bulk oil. (b) The phase-separated dispersions after the rinsing with water: the capsules formed at 6 and 9 mM KOleate are destroyed. (c) After the rinsing: micrographs of samples from the subphase (with common emulsion drops) and cream (with large drops formed after capsule destruction). The scaling caliper shows 20 μm.

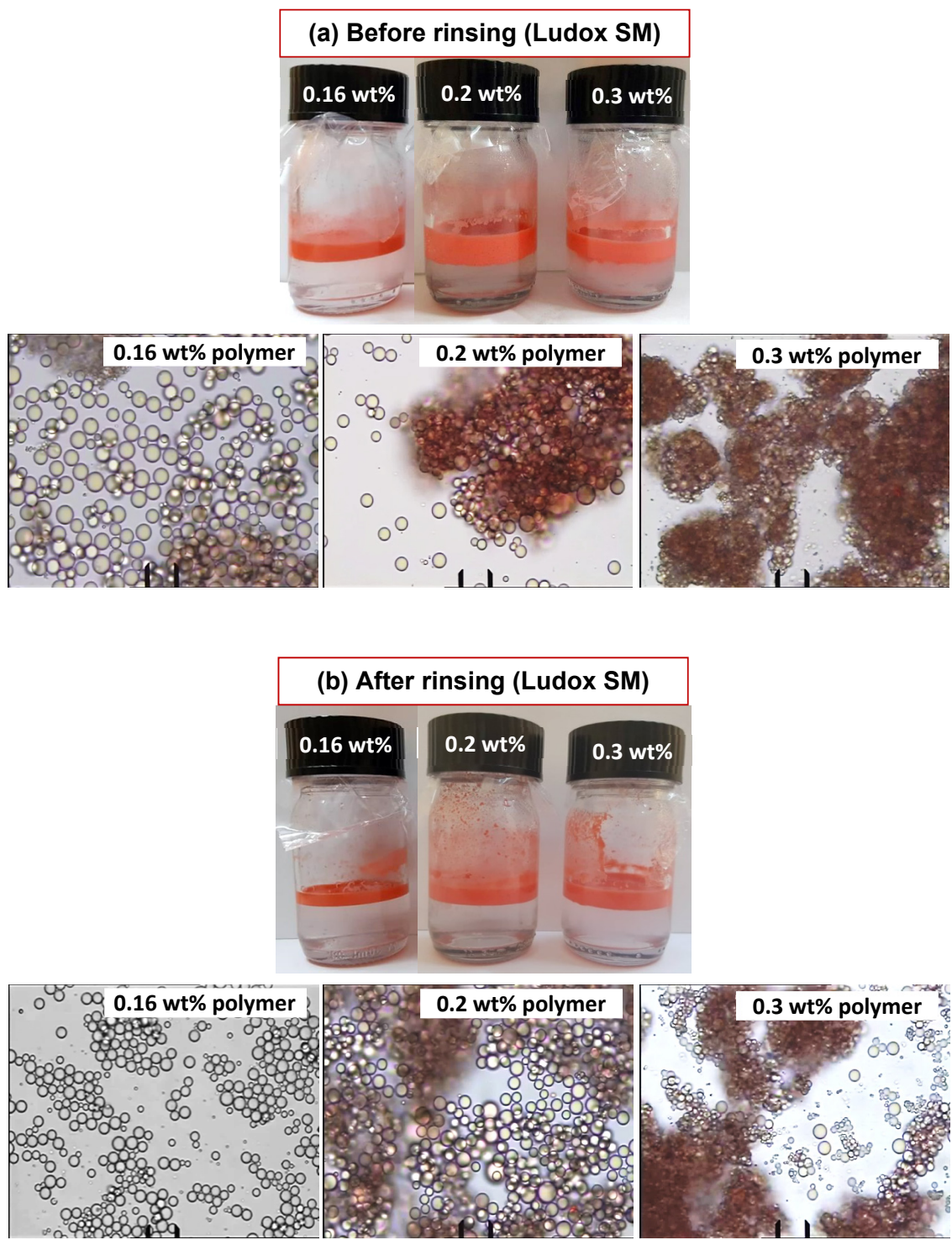

**Fig. S11.** <u>Ludox SM nanoparticles</u>. Comparison of the capsules before and after rinsing at three concentrations of the polymer Capbopol 971P denoted on the caps of the vials. The other ingredients are 0.1 wt% Ludox SM + 12 mM KOleate + 500 mM NaCl + 10 vol% limonene. Photographs and micrographs taken (a) before and (b) after the rinsing with water. The scaling caliper shows 20 μm. One sees that the excess polymer leads to aggregation of the capsules.

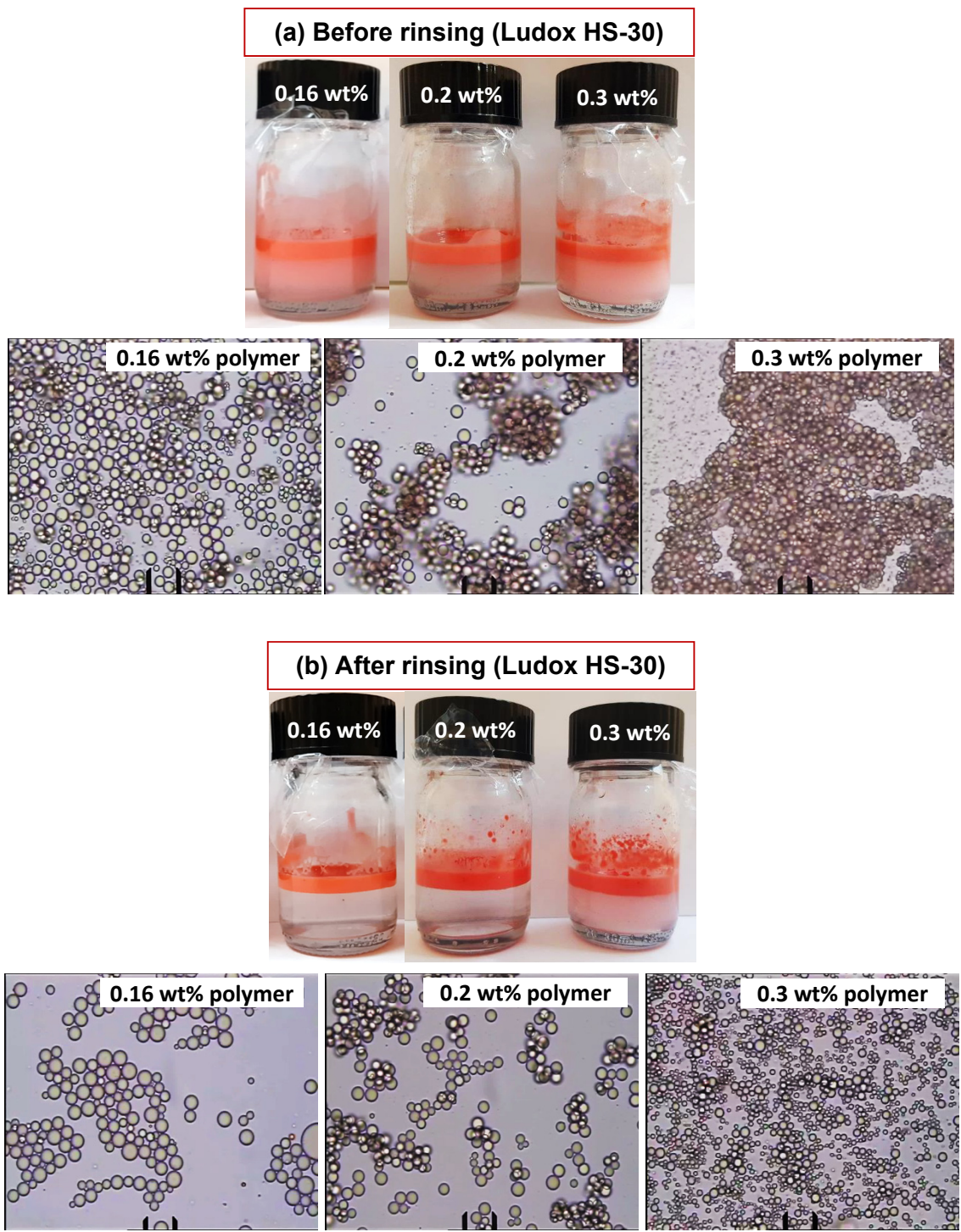

**Fig. S12.** <u>Ludox HS-30 nanoparticles</u>. Comparison of the capsules before and after rinsing at three concentrations of the polymer Capbopol 971P denoted on the caps of the vials. The other ingredients are 0.1 wt% Ludox HS-30 + 12 mM KOleate + 500 mM NaCl + 10 vol% limonene. Photographs and micrographs taken (a) before and (b) after the rinsing with water. The scaling caliper shows 20 μm. One sees that the excess polymer leads to the formation of smaller capsules of greater polydispersity, which can be explained with intercalation of polymer segments in between the adsorbed nanoparticles; see the main text.

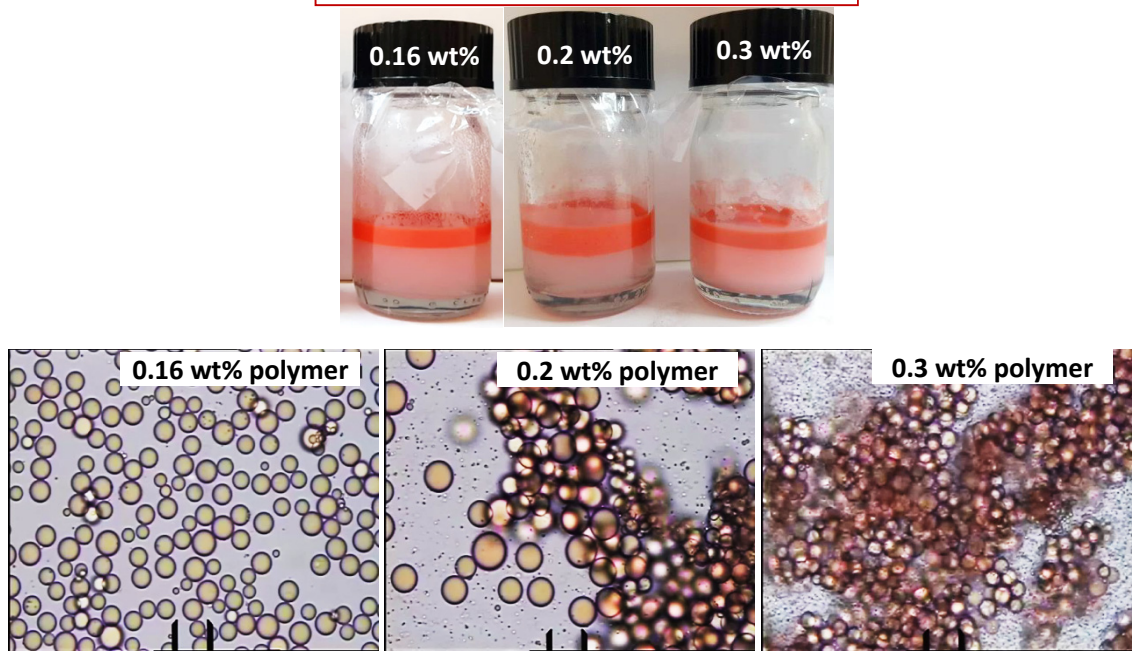
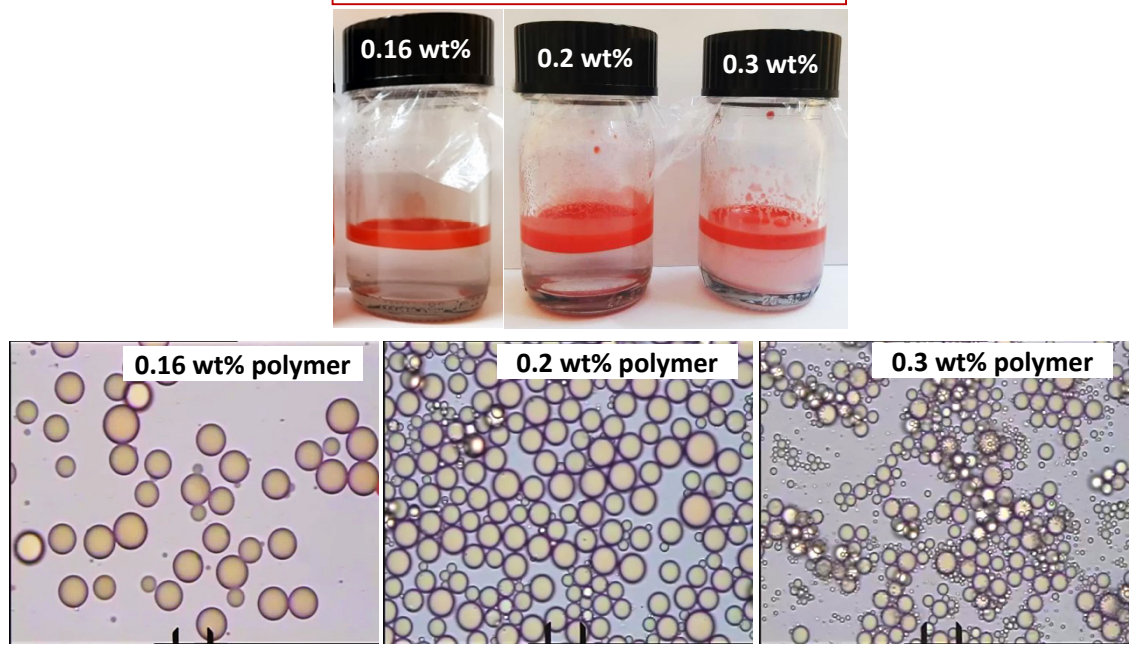

**Fig. S13.** <u>Ludox TM-50 nanoparticles</u>. Comparison of the capsules before and after rinsing at three concentrations of the polymer Capbopol 971P denoted on the caps of the vials. The other ingredients are 0.1 wt% Ludox TM-50 + 12 mM KOleate + 500 mM NaCl + 10 vol% limonene. Photographs and micrographs taken (a) before and (b) after the rinsing with water. The scaling caliper shows 20 μm. One sees that the excess polymer leads to the formation of smaller capsules of markedly greater polydispersity, which can be explained with intercalation of polymer segments in between the adsorbed nanoparticles; see the main text.

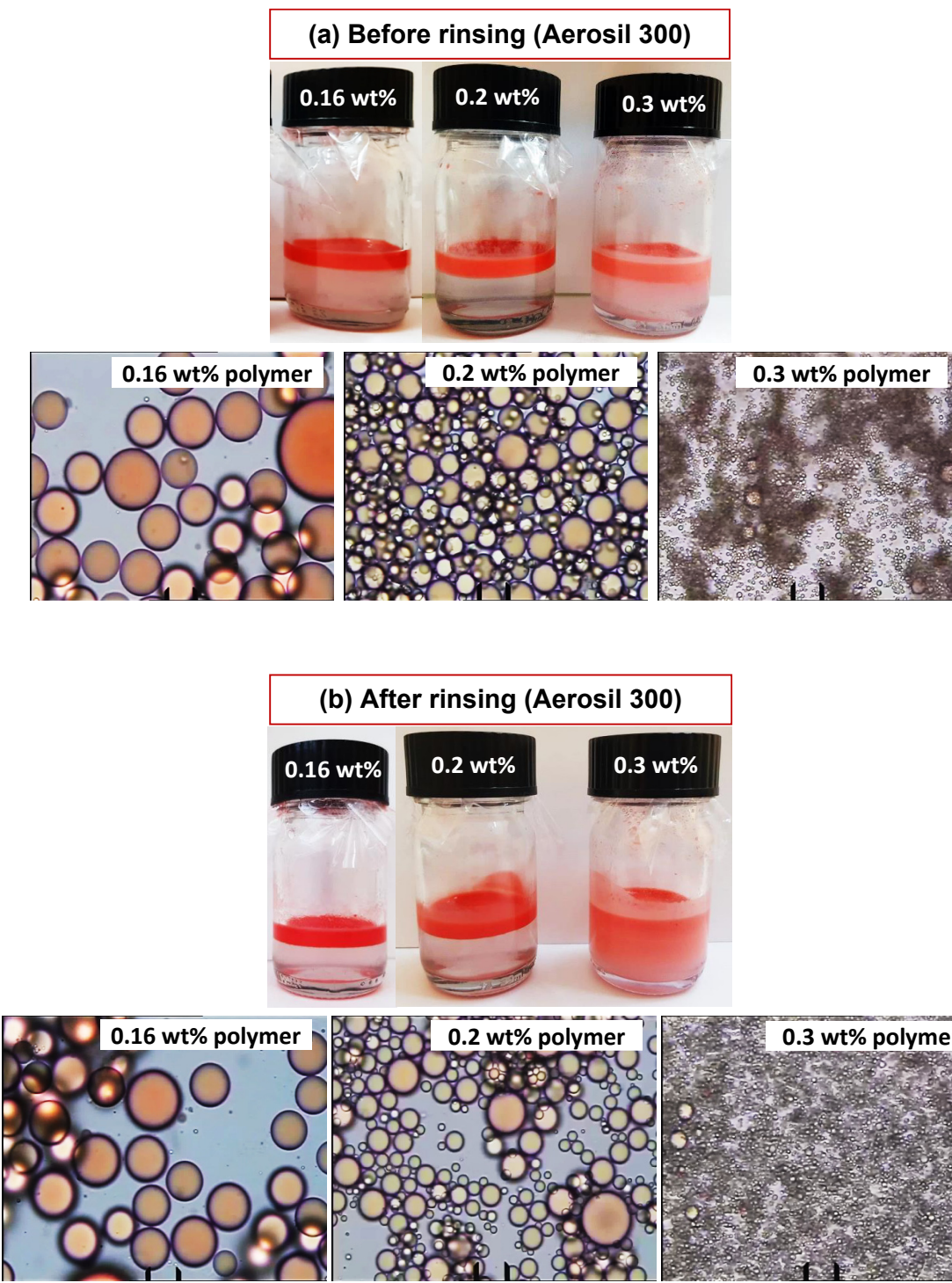

**Fig. S14.** <u>Aerosil 300 nanoparticles</u>. Comparison of the capsules before and after rinsing at three concentrations of the polymer Capbopol 971P denoted on the caps of the vials. The other ingredients are 0.1 wt% Aerosil 300 + 12 mM KOleate + 500 mM NaCl + 10 vol% limonene. Photographs and micrographs taken (a) before and (b) after the rinsing with water. The scaling caliper shows 20 μm. One sees that the excess polymer leads to the formation of smaller capsules of much higher polydispersity, which can be explained with intercalation of polymer segments in between the adsorbed nanoparticles; see the main text.

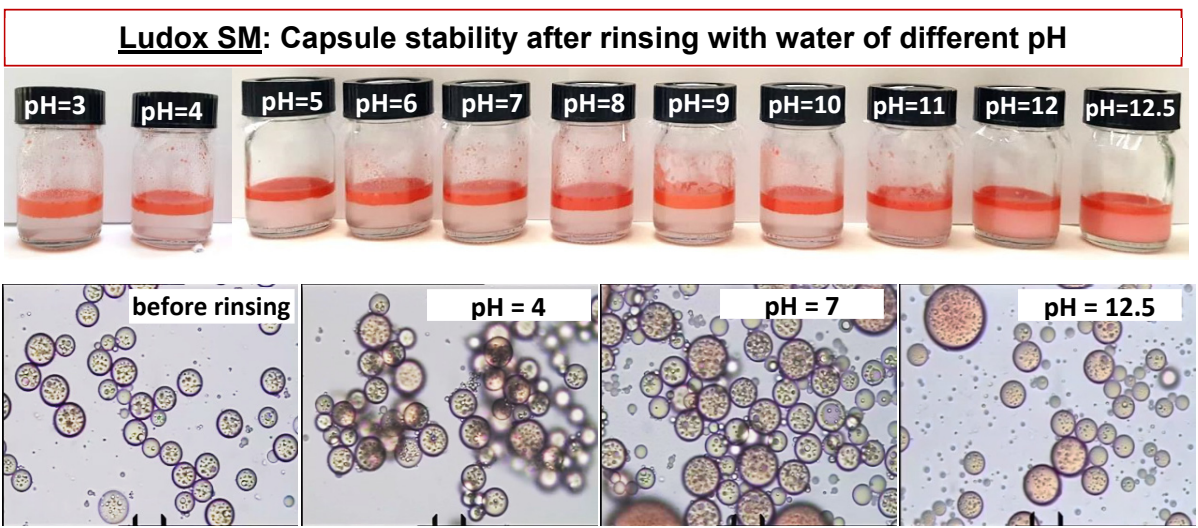

**Fig. S15.** Ludox SM nanoparticles: Stability of the capsules upon rinsing with water of different pH values denoted on the caps of the vials – photographs of vials with the phase separated dispersion and micrographs of samples from the cream (the scaling caliper shows 20 μm). The used composition is 0.1 wt% Ludox SM + 12 mM KOleate + 0.16 wt% Carbopol 971P + 500 mM NaCl + 10 vol% SFO. Subphase of red color indicates capsule destruction. The small water drops in the oil (double emulsion) are stabilized by an admixture of oleic acid in the SFO [1].

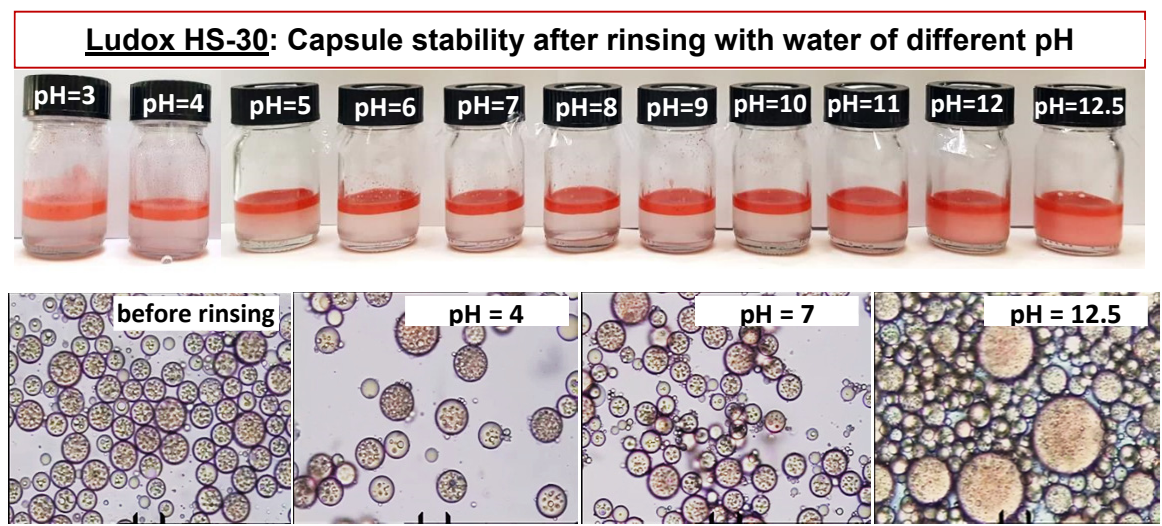

**Fig. S16.** Ludox HS-30 nanoparticles: Stability of the capsules upon rinsing with water of different pH values denoted on the caps of the vials – photographs of vials with the phase separated dispersion and micrographs of samples from the cream (the scaling caliper shows 20 μm). The used composition is 0.1 wt% Ludox HS-30 + 12 mM KOleate + 0.16 wt% Carbopol 971P + 500 mM NaCl + 10 vol% SFO. Subphase of red color indicates capsule destruction. The small water drops in the oil (double emulsion) are stabilized by an admixture of oleic acid in the SFO [1].

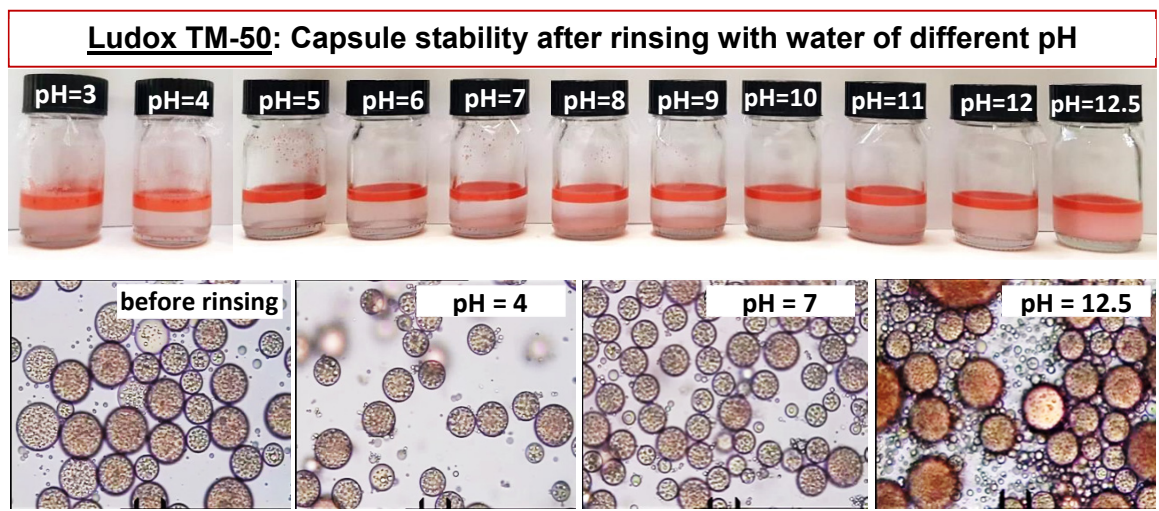

**Fig. S17.** <u>Ludox TM-50 nanoparticles</u>: Stability of the capsules upon rinsing with water of different pH values denoted on the caps of the vials – photographs of vials with the phase separated dispersion and micrographs of samples from the cream (the scaling caliper shows 20 μm). The used composition is 0.1 wt% Ludox TM-50 + 12 mM KOleate + 0.16 wt% Carbopol 971P + 500 mM NaCl + 10 vol% SFO. Subphase of red color indicates capsule destruction. The small water drops in the oil (double emulsion) are stabilized by an admixture of oleic acid in the SFO [1].

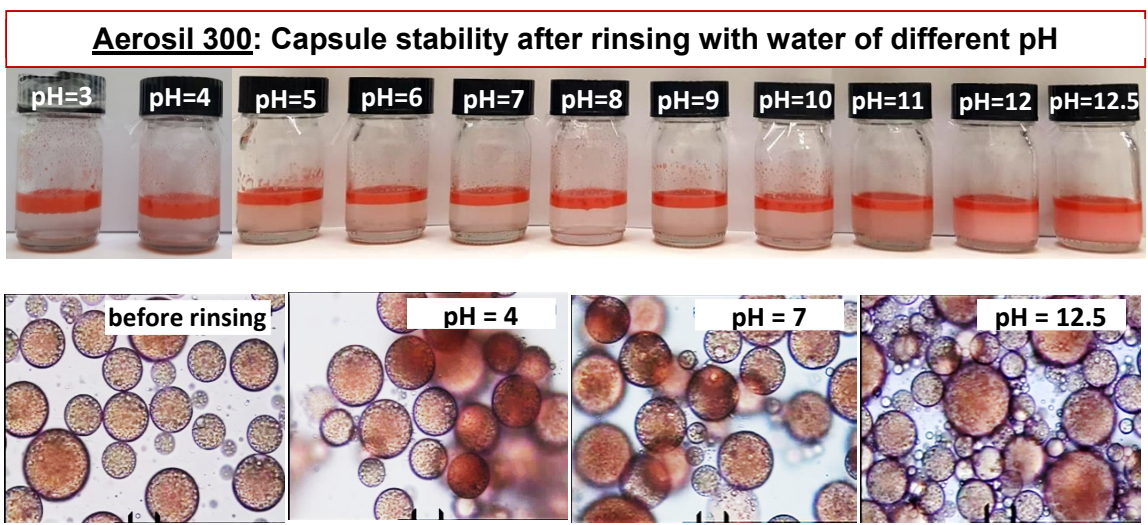

**Fig. S18.** <u>Aerosil-300 nanoparticles</u>: Stability of the capsules upon rinsing with water of different pH values denoted on the caps of the vials – photographs of vials with the phase separated dispersion and micrographs of samples from the cream (the scaling caliper shows 20 μm). The used composition is 0.1 wt% Aerosil 300 + 12 mM KOleate + 0.16 wt% Carbopol 971P + 500 mM NaCl + 10 vol% SFO. Subphase of red color indicates capsule destruction. The small water drops in the oil (double emulsion) are stabilized by an admixture of oleic acid in the SFO [1].